\documentclass{article}

\usepackage{arxiv}

\usepackage[utf8]{inputenc} 
\usepackage[T1]{fontenc}    
\usepackage{hyperref}       
\usepackage{url}            
\usepackage{booktabs}       
\usepackage{amsfonts}       
\usepackage{nicefrac}       
\usepackage{microtype}      
\usepackage{graphicx}
\usepackage{doi}

\usepackage{amsmath}
\usepackage{subcaption}
\usepackage{multirow}
\usepackage{xcolor}

\title{Sandwiched Hybrid Waveguide Platform for Integrated Photonics Application}

\author{ \href{https://orcid.org/0000-0003-3882-9845}{\includegraphics[scale=0.06]{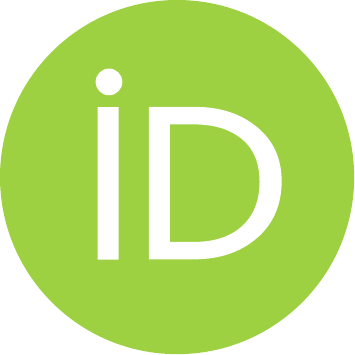}\hspace{1mm}Rahul K Dash} \\
	Centre for Nanoscience and Engineering\\
	Indian Institute of Science\\
	Bangalore, 560012 \\
	\texttt{rahuld@iisc.ac.in} \\
	\And
	\href{https://orcid.org/0000-0003-2670-7058}{\includegraphics[scale=0.06]{orcid.pdf}\hspace{1mm}Shankar Kumar Selvaraja} \\
	Centre for Nanoscience and Engineering\\
	Indian Institute of Science\\
	Bangalore, 560012 \\
	\texttt{shankarks@iisc.ac.in} \\
}

\begin{document}
\maketitle

\begin{abstract}
	We propose and demonstrate a hybrid waveguide platform using layered amorphous silicon and silicon nitride. The waveguide offers more degrees of freedom to design waveguides with desired confinement, effective index and polarization birefringence. Unlike single core material, the proposed waveguide offers design flexibility, and light confinement in the layers is polarization-dependent. We present a detailed waveguide design and analysis of efficient fiber-chip grating couplers with a coupling efficiency of -3.27 dB and -8 dB for $TE$ and $TM$ polarization, respectively. The couplers offer a 3dB bandwidth of 100 nm. Furthermore, we demonstrate excitation of TE and TM modes exploiting the polarization-dependent confinement using thermo-optic characteristics of a ring resonator.
\end{abstract}


\section{Introduction}
\begin{figure}[b!]

\begin{subfigure}{0.5\textwidth}
\centering\includegraphics[width=6 cm]{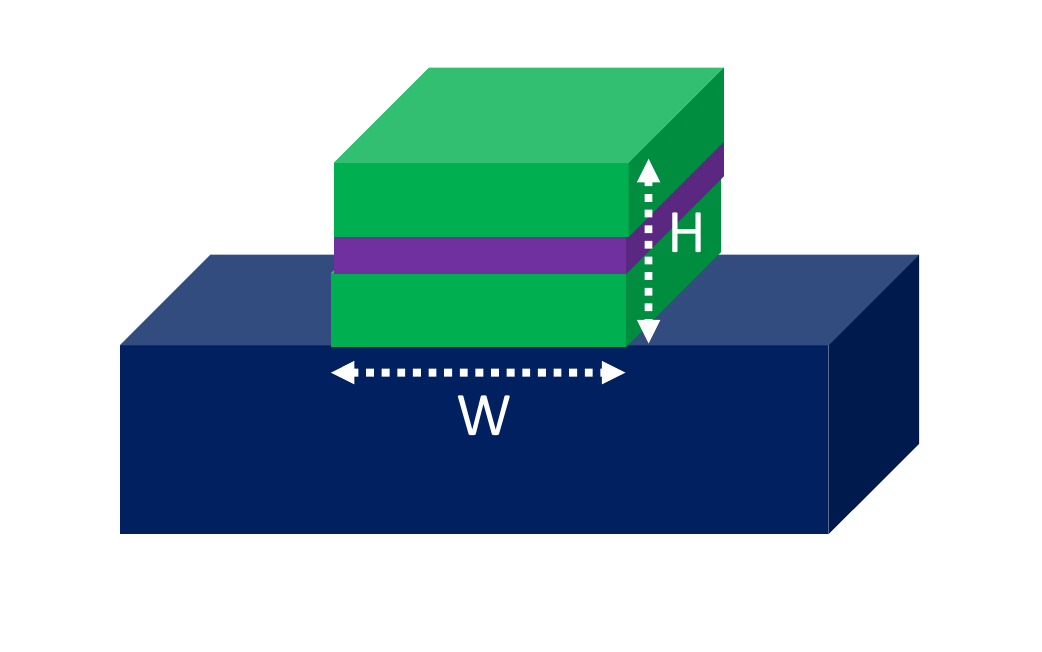}
\centering\caption{3D view}
\label{fig:sim11}
\end{subfigure}
\begin{subfigure}{0.5\textwidth}
\includegraphics[width=6cm]{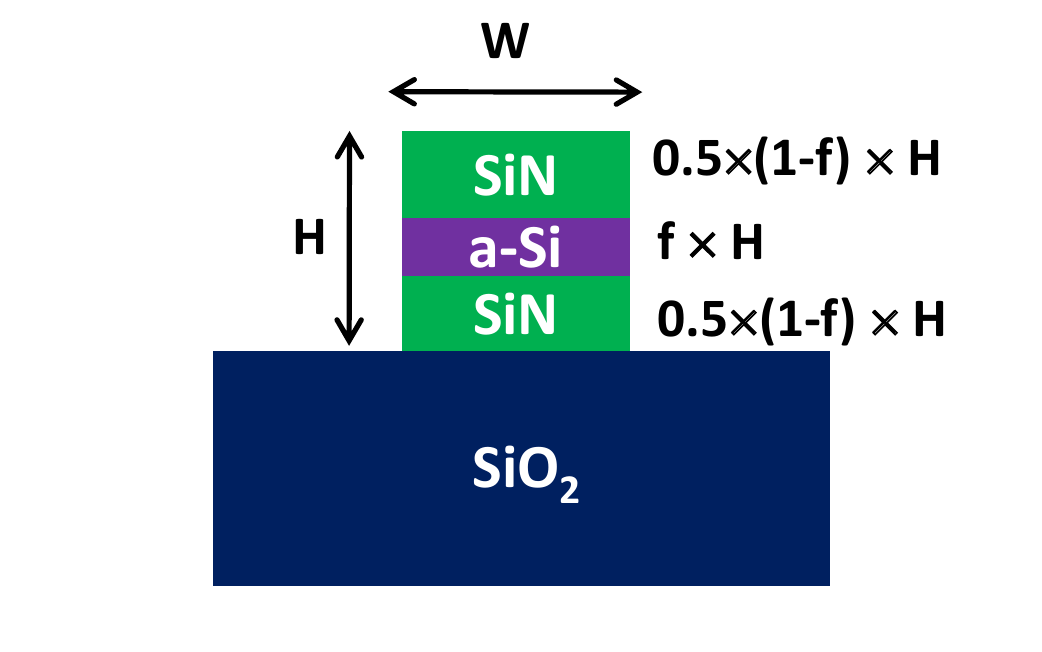}
\caption{Cross sectional view}
\label{fig:sim12}
\end{subfigure}
\caption{ Schematic of the hybrid sandwich waveguide (a) 3D view and (b) Cross-sectional view showing various design parameters of waveguide.}
\label{fig:im1}
\end{figure}

Photonic integrated circuits (PIC) offer a versatile platform for on-chip manipulation of light for various linear and nonlinear optical signal processing. The optical waveguide forms an essential part of the photonic circuit through which optical signals are routed. In addition, waveguides also forms the basis for various waveguide components. Conventional waveguide structures such as wire/strip, rib, and slot are formed with a core made of single uniform refractive index material that dictates the overall properties \cite{Bogaerts2011,Almeida2004}. A core with a uniform refractive index limits the waveguide properties by restricting the effective index and polarization birefringence. These restrictions can be addressed by appropriately designing the refractive index profile of the core. A graded index waveguide was reported to have a broadband response with minimized losses in the mid-infrared (MIR) region \cite{Brun2014}. Another reported structure is augmented low-index guide (ALIG) with a low-index material deposited on a high index material that aims to confine light in the low index medium  \cite{Alam2017}. The core of ALIG, however, has inherent asymmetry that can result in undesirable mode conversions in waveguide taper \cite{Dai2012,Vermeulen2010}. Triplex is another waveguide platform that offers low optical loss and desirable circuit element implementation \cite{triplex}. Thus, a waveguide with a non-uniform index profile provides some advantages and has exotic properties which need more investigation. 

In this work, we propose and demonstrate a symmetric stack where a central high index material is sandwiched between two medium index materials. The high index material is chosen as amorphous silicon (a-Si), and the medium index material is chosen as silicon nitride ($SiN$). Both $a-Si$ and $SiN$  can be deposited by standard deposition techniques, and they are promising materials for PIC \cite{KumarSelvaraja2014,Selvaraja2009_1,Baets2016}. Figure \ref{fig:sim11} and \ref{fig:sim12} shows a schematic and the geometrical parameters in cross-sectional view of the proposed structure, respectively. The structure has three geometrical parameters that can be varied to tune the properties of the waveguide. Height ('H') and width ('W') are similar parameters as in the wire waveguide. But, this structure has an additional parameter, the relative thickness of high index material with respect to the total height. This makes the waveguide properties highly tunable. The relative thickness of high-index material and medium-index material can be replaced by a single variable, simplifying the analysis. The relative thickness of high index material will be referred to as fraction (`\textit{f}') from here on. The parameter fraction is defined as \textit{f} = Thickness of high index/Total height ('H'). The parameter `\textit{f}' provides a powerful way to analyze the system by having a single variable instead of two thickness variables. The implications of the fraction on the waveguide properties are discussed in detail for both air and oxide clad, followed by a demonstration of $TE$ and $TM$ gratings for fiber-to-chip coupling in the proposed structure. The measured coupling efficiency for $TE$ grating is -3.27 dB per coupler, and $TM$ grating is -8 dB per coupler. Finally, the excitation of a particular mode (TE/TM) is confirmed by thermo-optic measurements based on the spectral response of a ring resonator.

\section{Waveguide Design }\label{Waveguide_Design}

A wire waveguide can be designed by appropriately choosing the width and height to suit a particular application. The proposed waveguide can be further tuned by varying the fraction. This unique parameter provides flexibility in effective index tuning, confinement in different materials and polarization birefringence control which are restricted in uniform index waveguides. We consider waveguide design with top cladding as air and oxide. The height of the proposed waveguide is fixed at 500 nm to ensure mode confinement in the entire work.

\subsection{Waveguide design in air clad}

\begin{figure}[ht!]
\centering\includegraphics[width=15cm]{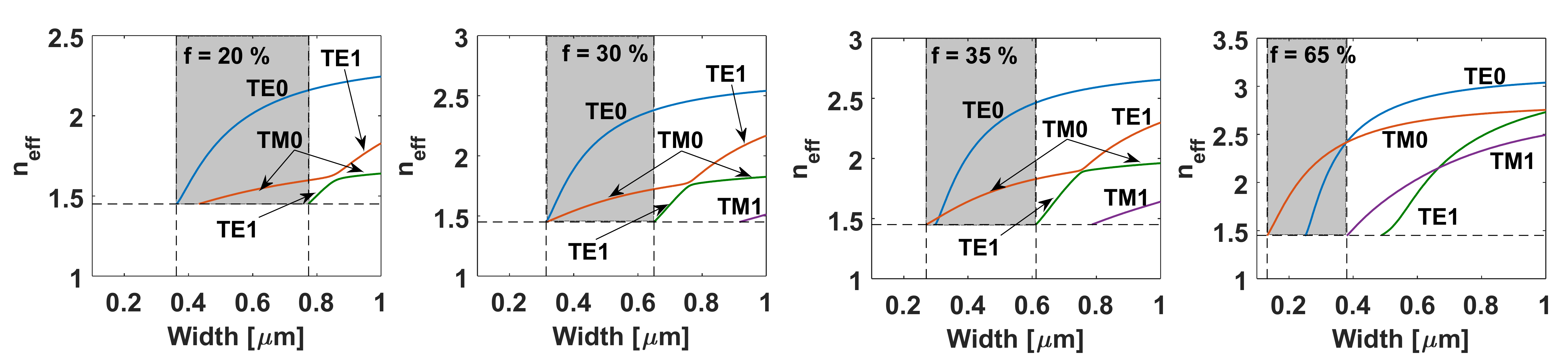}
\caption{Effective index for various fraction \textit{'f'} showing fundamental and higher order modes for air clad. The shaded region shows the single-mode regime.}
\label{im2}
\end{figure}

The waveguide design is performed using a numerical mode-solver. The refractive index of SiN is considered as 1.89 and a-Si as 3.54 and the wavelength of operation is 1550 nm. Simulations were performed to study the effect of fraction (\textit{f}) on the waveguide properties. The evolution of effective mode index as a function of width is found for various \textit{f}, and a few of them are shown in Figure \ref{im2}. In fig.\ref{im2}, the single-mode region is marked by the shaded area. For higher fraction \textit{f}, we observe propagating $TM_{0}$ with higher index than the $TE_{0}$ mode in the single-mode regime.

\begin{figure}
\begin{subfigure}{0.5\textwidth}
    \centering
    \includegraphics[width=5cm]{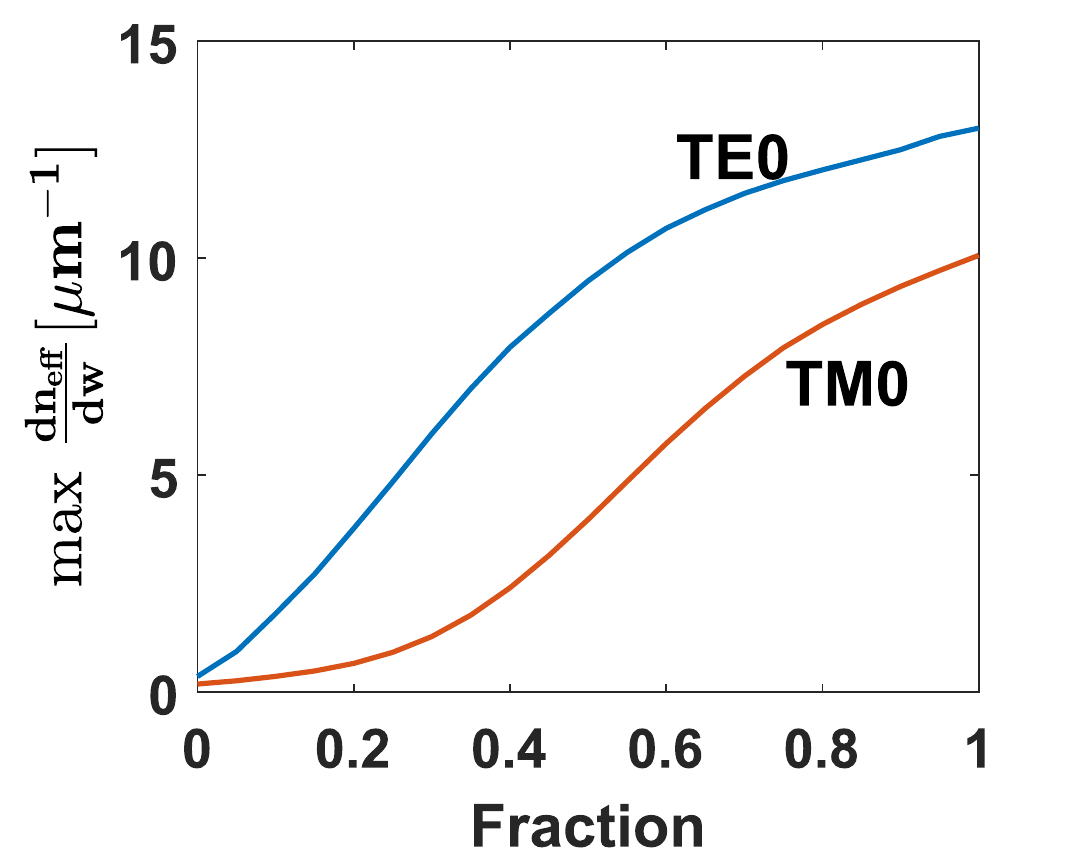}
    \centering\caption{}
    \label{fig:sim31}
\end{subfigure}
\begin{subfigure}{0.5\textwidth}
    \centering
    \includegraphics[width=5cm]{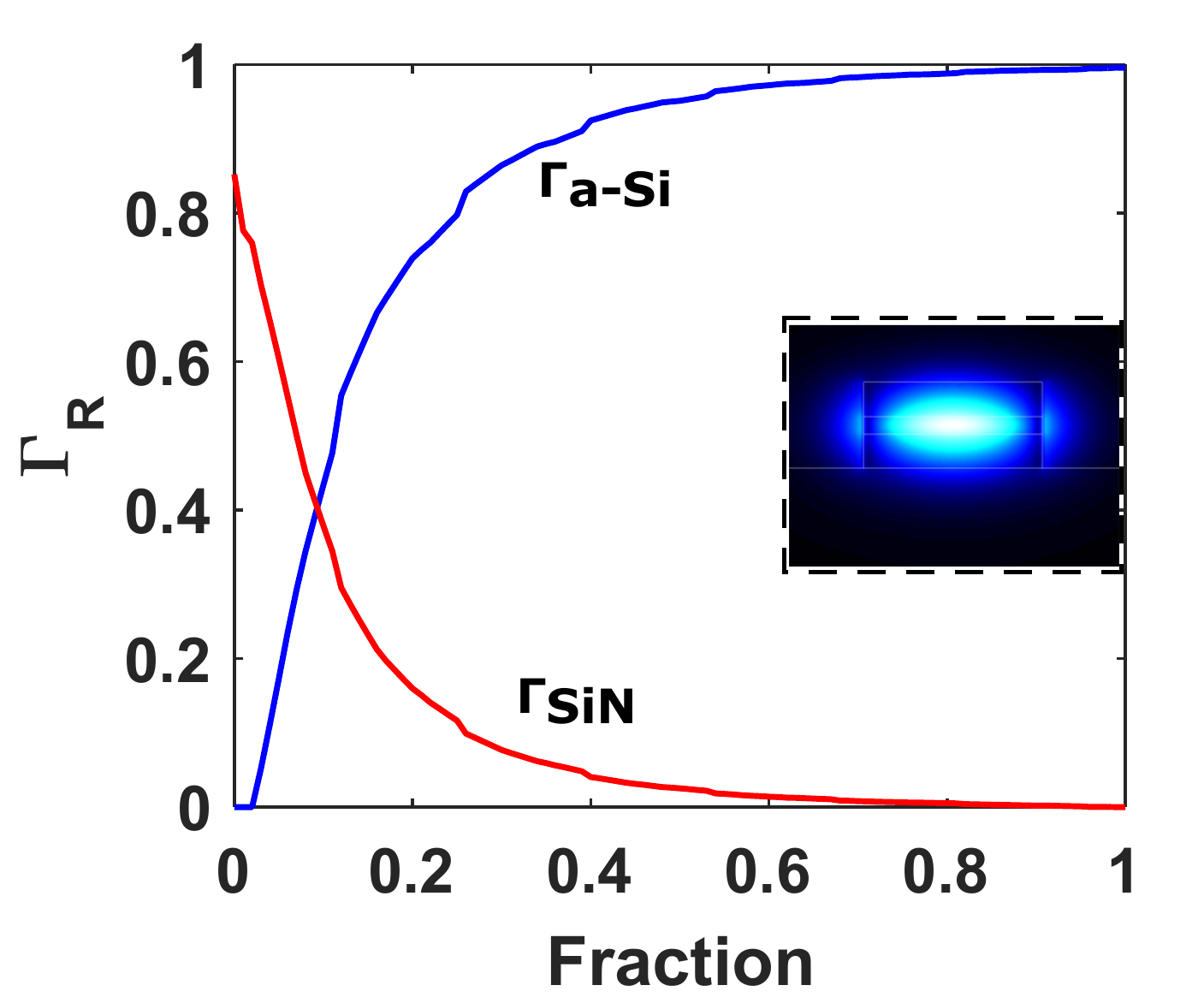}
    \centering\caption{$TE$ mode}
    \label{fig:sim32}
\end{subfigure}
\begin{subfigure}{0.5\textwidth}
    \centering
    \includegraphics[width=5cm]{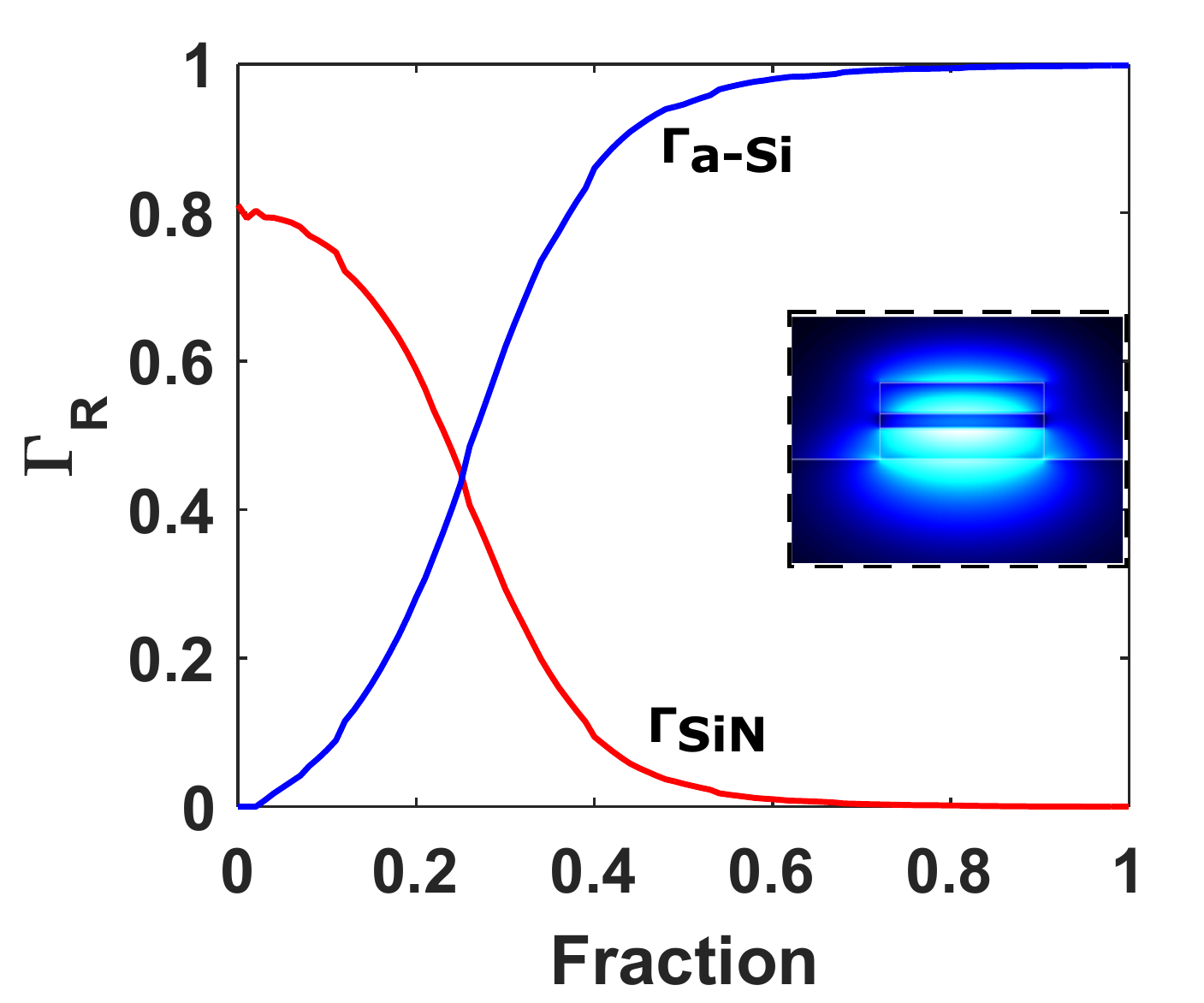}
    \centering\caption{$TM$ mode}
    \label{fig:sim33}
\end{subfigure}
    \begin{subfigure}{0.5\textwidth}
    \centering
    \includegraphics[width=5cm]{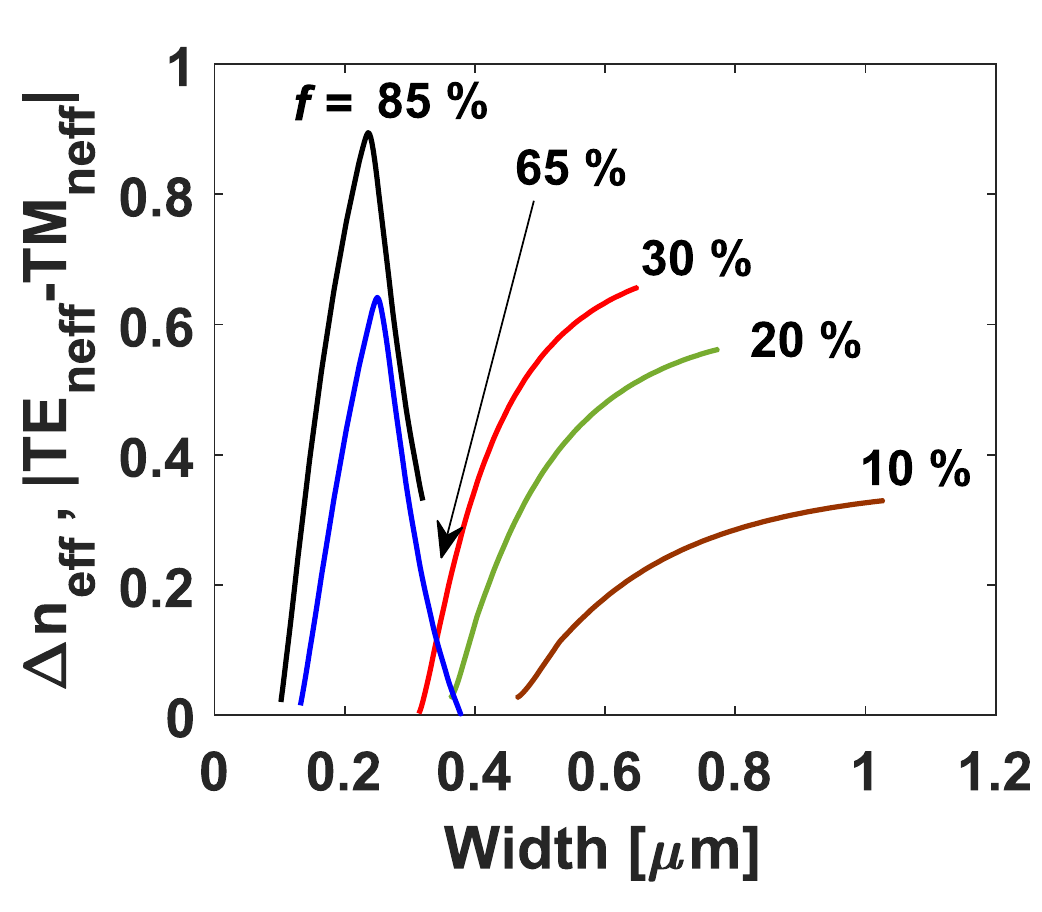}
    \centering\caption{}
\label{fig:sim34}
\end{subfigure}

\caption{ Analysis of air clad waveguide (a) Maximum slope of effective index vs width as a function of fraction. Confinement of (b)  $TE$ mode and (c) $TM$ mode in $a-Si$ and $SiN$ layers as a function of fraction. Inset shows the field distribution for f = 20\%.(d) Polarization birefringence vs width as a function of fraction.}
\label{fig:im3}
\end{figure}

The effective index is a fundamental property that determines the functional property of various waveguide devices. Variation in the effective index due to width and fraction will result in suboptimal device performance. Figure \ref{fig:sim31} shows the $max(dn_{eff}/dW)$ at different fractions \textit{f} in the single-mode regime. It can be clearly observed that the $TE$ mode is highly sensitive to width variation and \textit{f} variation than the $TM$ mode. This implies that the $TM$ modes are tolerant to layer-thickness and width variation compared to $TE$.

It is important to note that confinement of light is in both $a-Si$ and $SiN$ due to the hybrid structure of the waveguide. The waveguide will be henceforth referred to as hybrid waveguide. The hybrid nature of the waveguide demands calculation of confinement for both $a-Si$ and $SiN$ regions. The confinement factor in a region R is defined as \cite{Tien2010}: 

\begin{equation}\label{eq2}
\Gamma_R = \frac{\iint_R ((I(x,y))^2 \,dx\,dy}{\iint_\infty (I(x,y))^2 \,dx\,dy}
\end{equation}

The confinement in the waveguide depends on the polarization. Figure \ref{fig:sim32} and \ref{fig:sim33} show the confinement in the $a-Si$ and $SiN$ layer for a 700 nm wide waveguide to have a broad range of fraction values where the width lies in the single mode region. It can be clearly observed that for $TE$ mode the light confinement is higher in the $a-Si$ layer for \textit{f} $>$ 10 \% and for $TM$ mode it is higher for \textit{f} $>$ 25 \%. Thus, the factor \textit{f} can be tuned to have desired confinement in either material to best suit any application. An interesting observation is that the confinement of light for a given geometry in $a-Si$ and $SiN$ layers is polarization-dependent. Light can be confined into a higher index with $TE$ mode and into a medium index with $TM$ mode. It can be concluded from Figures \ref{fig:sim32} and \ref{fig:sim33} that a fraction of 20 \% is ideal for confining light into $a-Si$ with $TE$ mode and into $SiN$ with $TM$ mode. This typical behaviour can be explained by the boundary conditions of Maxwell’s equation. The normal component of displacement density should be continuous, making the electric field in the $SiN$ layer for $TM$ mode as
\begin{equation}\label{eq1}
E_{SiN} = (\frac{n_{a-Si}}{{n_{SiN}}})^2\times E_{a-Si}
\end{equation} 

In some applications, such as wavelength division multiplexing (WDM) a low polarization sensitivity \cite{Trinh1997} is desired. However, in some cases, high polarization extinction can be used for creating polarization-selective phase delay\cite{Yang2008}. Thus, a flexible control of birefringence ($dn_{eff}=TEn_{eff}-TMn_{eff}$) is highly desirable. In the hybrid platform, birefringence can be controlled by adjusting ‘f’. Figure \ref{fig:sim34} shows the birefringence for various fractions in the single mode region. The slope of birefringence vs width continues to increase with fraction. Thus, the trend seems to push the birefringence to higher values. But, it is observed that between f = 35 \% to f = 65\% , there exists an inflection point where the birefringence becomes negligible. This can be exploited to suit the particular application.

\subsection{Waveguide design in oxide clad}
\begin{figure}[ht!]
\centering\includegraphics[width=15cm]{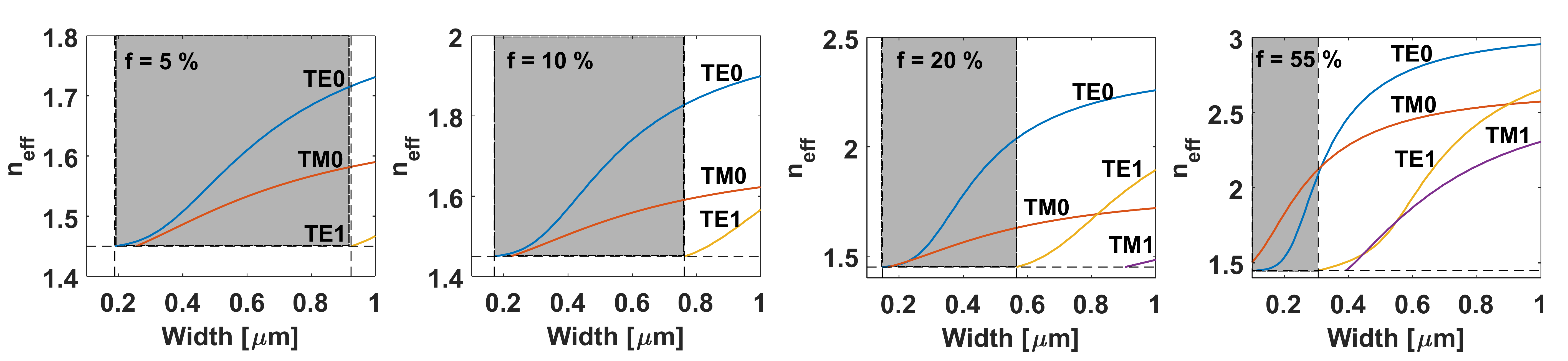}
\caption{Effective index for various fraction \textit{f} showing fundamental and higher order modes for oxide clad. The shaded region shows the single-mode regime.}
\label{im4}
\end{figure}

\begin{figure}[ht!]
\begin{subfigure}{0.5\textwidth}
\centering\includegraphics[width=5cm]{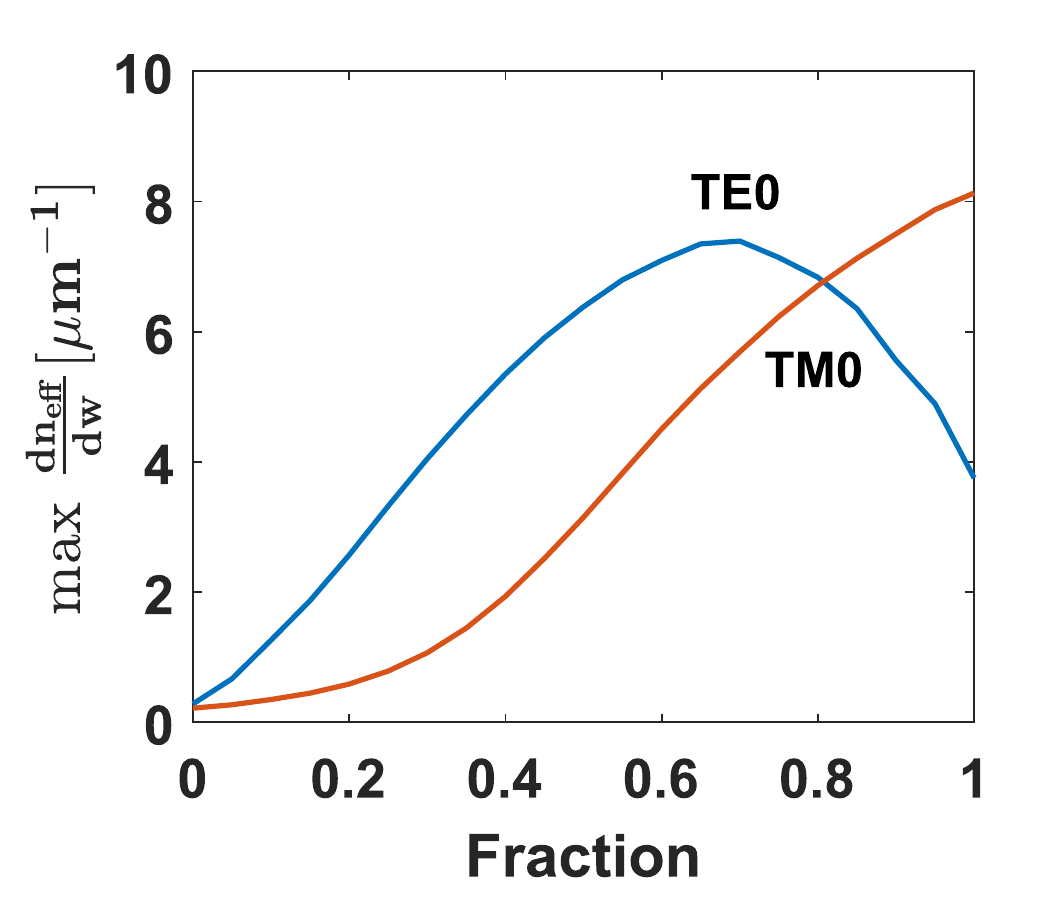}
\centering\caption{}
\label{subim51}
\end{subfigure}
\begin{subfigure}{0.5\textwidth}
\centering\includegraphics[width=5cm]{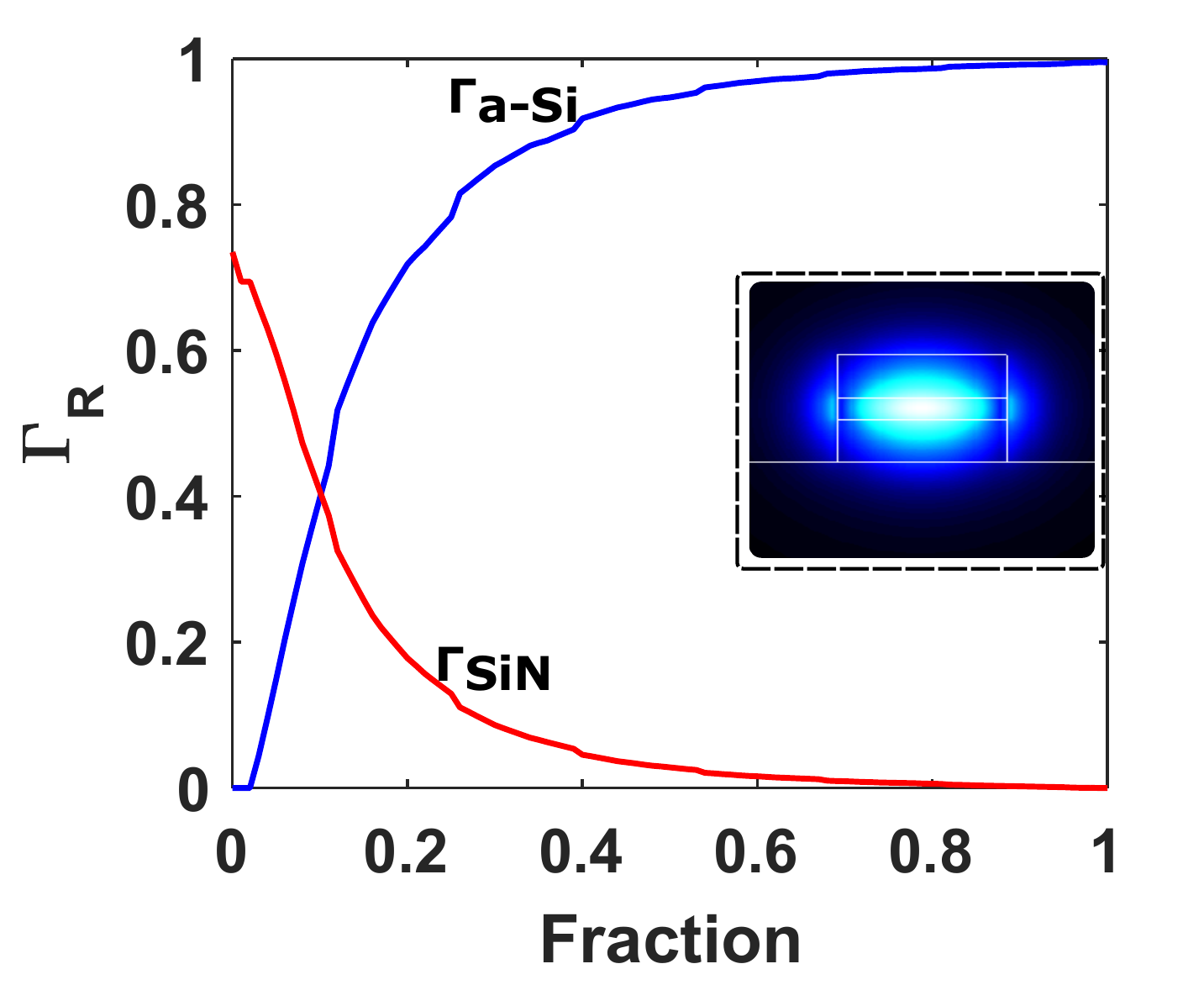}
\centering\caption{$TE$ mode}
\label{subim52}
\end{subfigure}
\begin{subfigure}{0.5\textwidth}
\centering\includegraphics[width=5cm]{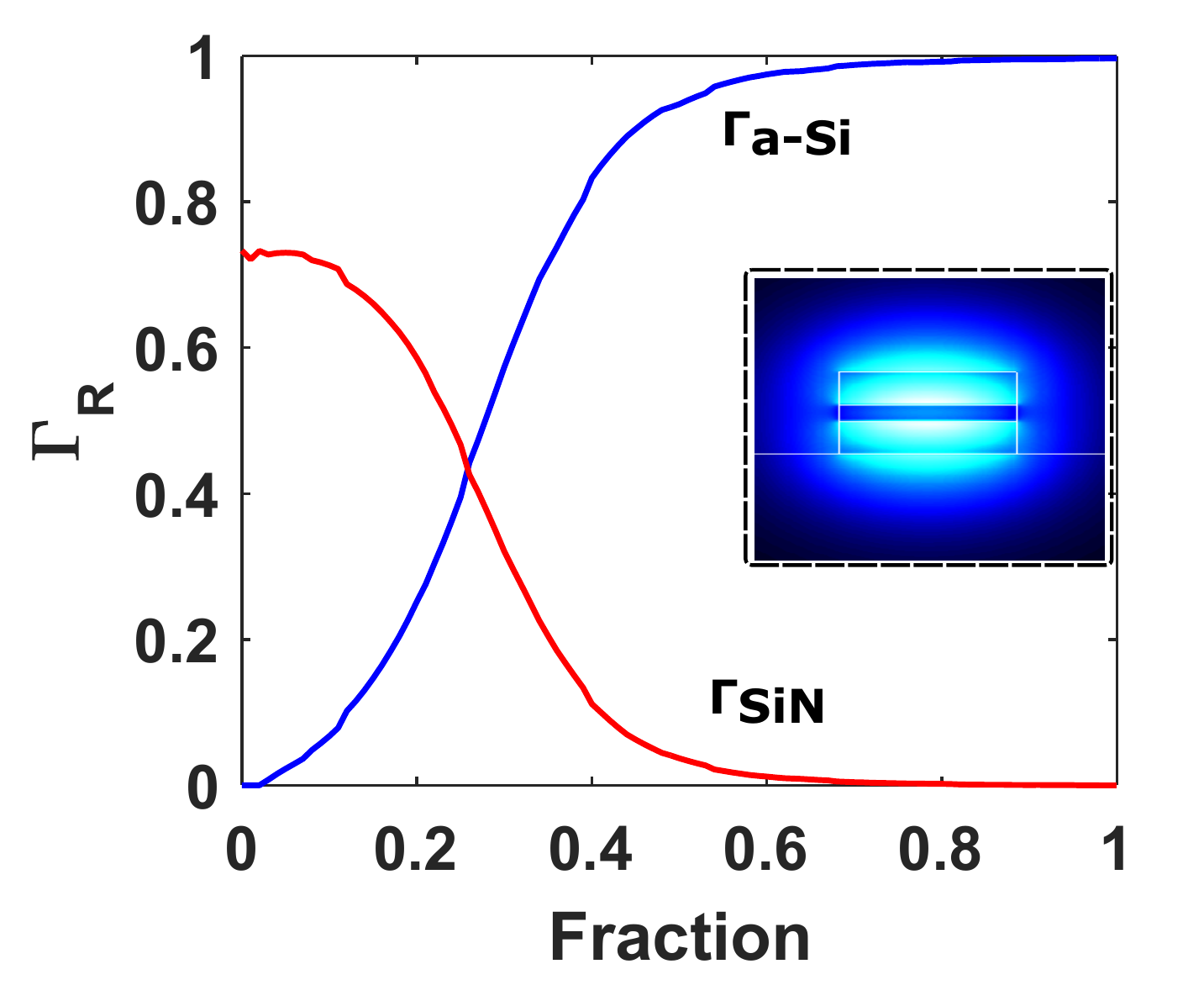}
\centering\caption{$TM$ mode}
\label{subim53}
\end{subfigure}
\begin{subfigure}{0.5\textwidth}
\centering\includegraphics[width=5cm]{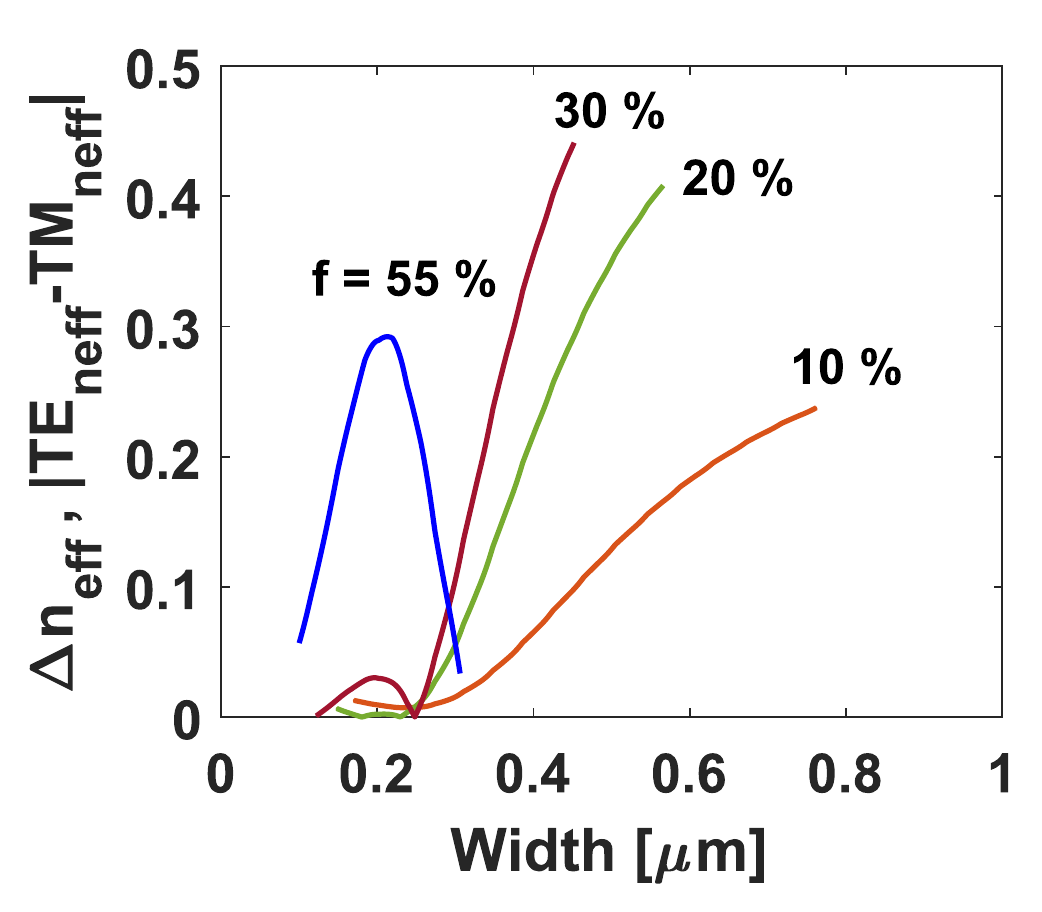}
\centering\caption{}
\label{subim54}
\end{subfigure}

\caption{ Analysis of oxide clad waveguide. (a) Maximum slope of effective index vs width as a function of fraction. Confinement of (b) $TE$ mode anbd (c) TM mode in $a-Si$ and $SiN$ layers as a function of fraction. Inset shows the field for $TE$ mode for f = 20\%. (d) Polarization birefringence vs width as a function of fraction.}
\label{fig:image2}
\end{figure}

The simulations of sandwich waveguide with oxide clad are performed to study the impact of fraction \textit{f} on the waveguide properties. The reduction in the refractive index contrast shifts the single-mode regime towards lower waveguide widths (Fig. \ref{im4}). For air clad, it is observed that there is anti-crossing between modes of different order indicating the existence of a hybrid mode that can result in undesirable mode conversions. Since the index-contrast is reduced, the $max(dn_{eff}/dW)$ is lowered, which implies that the waveguides are tolerant to dimensional variations, unlike air-clad.  

The confinement factor is calculated with the same procedure as for air clad. Figures \ref{subim52} and \ref{subim53} show confinement factor for $TE$ and $TM$ mode with oxide clad. The width is fixed at 500 nm. The trend is very similar to air clad. In both air-clad and oxide clad, a 20 \% fraction is suitable for light confinement. Figure \ref{subim54} shows the birefringence for oxide clad. The sensitivity to polarization for oxide clad is less than air clad. For larger values of fraction, the peak is flatter compared to air clad, giving a width tolerant birefringence.

\section{Fiber-chip coupling}
\subsection{Design and fabrication of $TE$ and $TM$ surface grating couplers}

Surface grating fiber-chip couplers are a versatile fiber-chip coupling scheme \cite{Nambiar2018}. A grating fiber-chip coupler for $TE_0$ and $TM_0$ is designed and developed to demonstrate the proposed waveguide in the 1550 nm wavelength band. The height in all the designs is considered as 500 nm and \textit{f} as 20 \%. The resulting thickness for the top and bottom $SiN$ layers is 200 nm, and $a-Si$ is 100 nm. For efficient coupling into $TE$ mode, the $a-Si$ layer is etched, and for $TM$ mode entire stack is etched to pattern the gratings. The designs are aimed at optimum diffraction to facilitate coupling into a given mode. The coupling efficiency (CE) is optimized by varying period, angle, duty-cycle, etch depth (only for $TE$ mode) and bottom oxide thickness by performing 2D finite difference time domain (FDTD) simulations. Particle swarm optimization (PSO) is used to optimize the coupling efficiency instead of parametric sweep due to a large number of parameters\cite{pso}. The dependence of various parameters is discussed in detail in the results section. The simulation results are summarized in table \ref{table:table1}.

\begin{table}[ht!]
\begin{center}
\caption{Performance summary of optimal grating coupler design.}
\begin{tabular}{|c||c|c|}

    \hline
     Parameter [units] & $TE$ & $TM$ \\
     \hline
   
     $n_{eff}$ & 1.94 & 1.58  \\
     Etched layer for grating &	$a-Si$ layer &	Full etch\\
    Coupling Efficiency [dB] & -2.52 & -3.9\\
	1 dB bandwidth [nm] &	57.6 &	54.4\\
    3 dB bandwidth [nm] &	98 &	94\\

     \hline

\end{tabular}   
\label{table:table1}
\end{center}
\end{table}

\begin{figure}[ht!]
\centering\includegraphics[width=15cm]{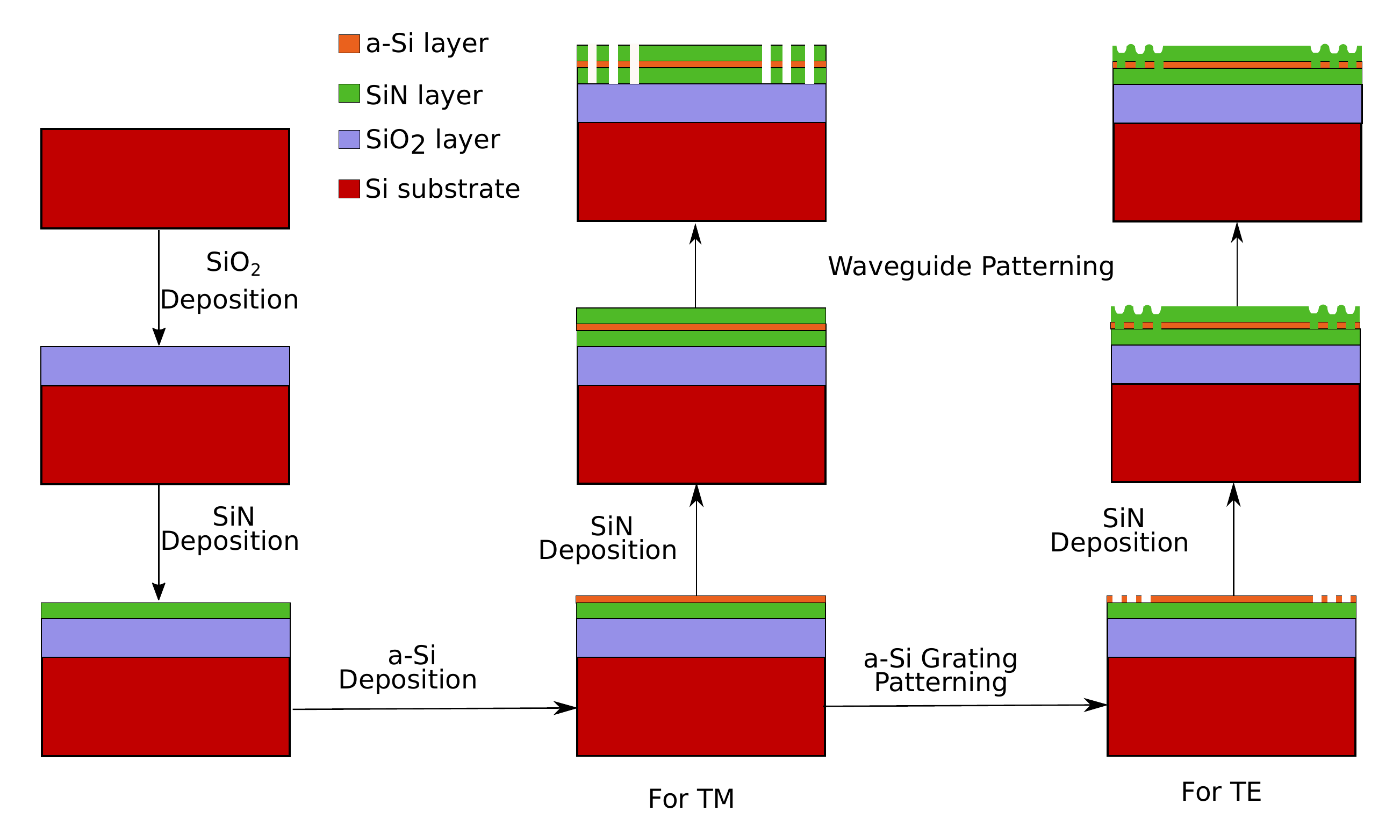}
\caption{Overview of $TE$ and $TM$ grating and  waveguide fabrication process flow.}
\label{fig:im6}
\end{figure}

\begin{figure}[t!]
\begin{subfigure}{0.33\textwidth}
\centering\includegraphics[width=4.3cm]{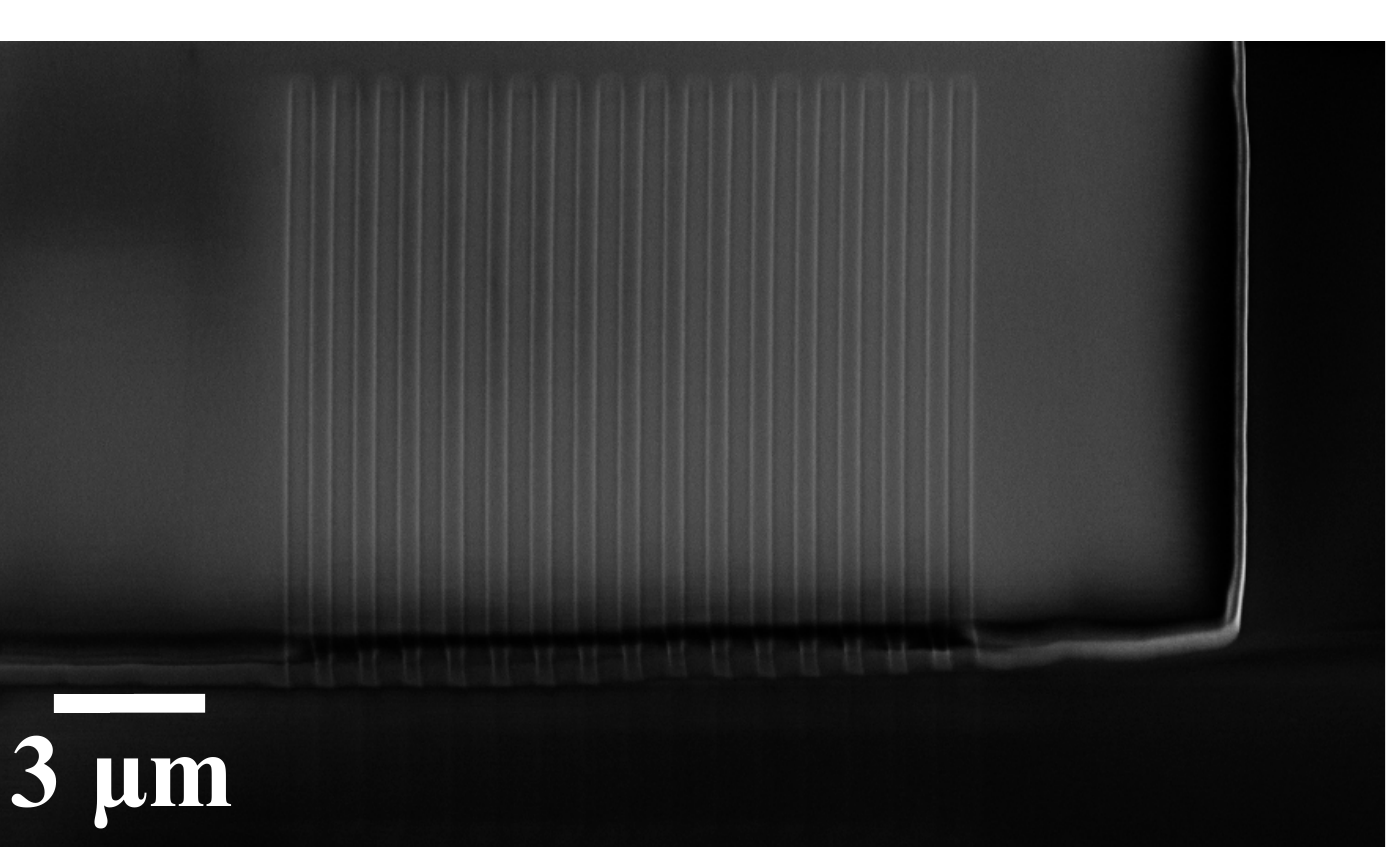}
\caption{$TE$ grating}
\label{fig:subim71}
\end{subfigure}
\begin{subfigure}{0.32\textwidth}
\centering\includegraphics[width=4.3cm]{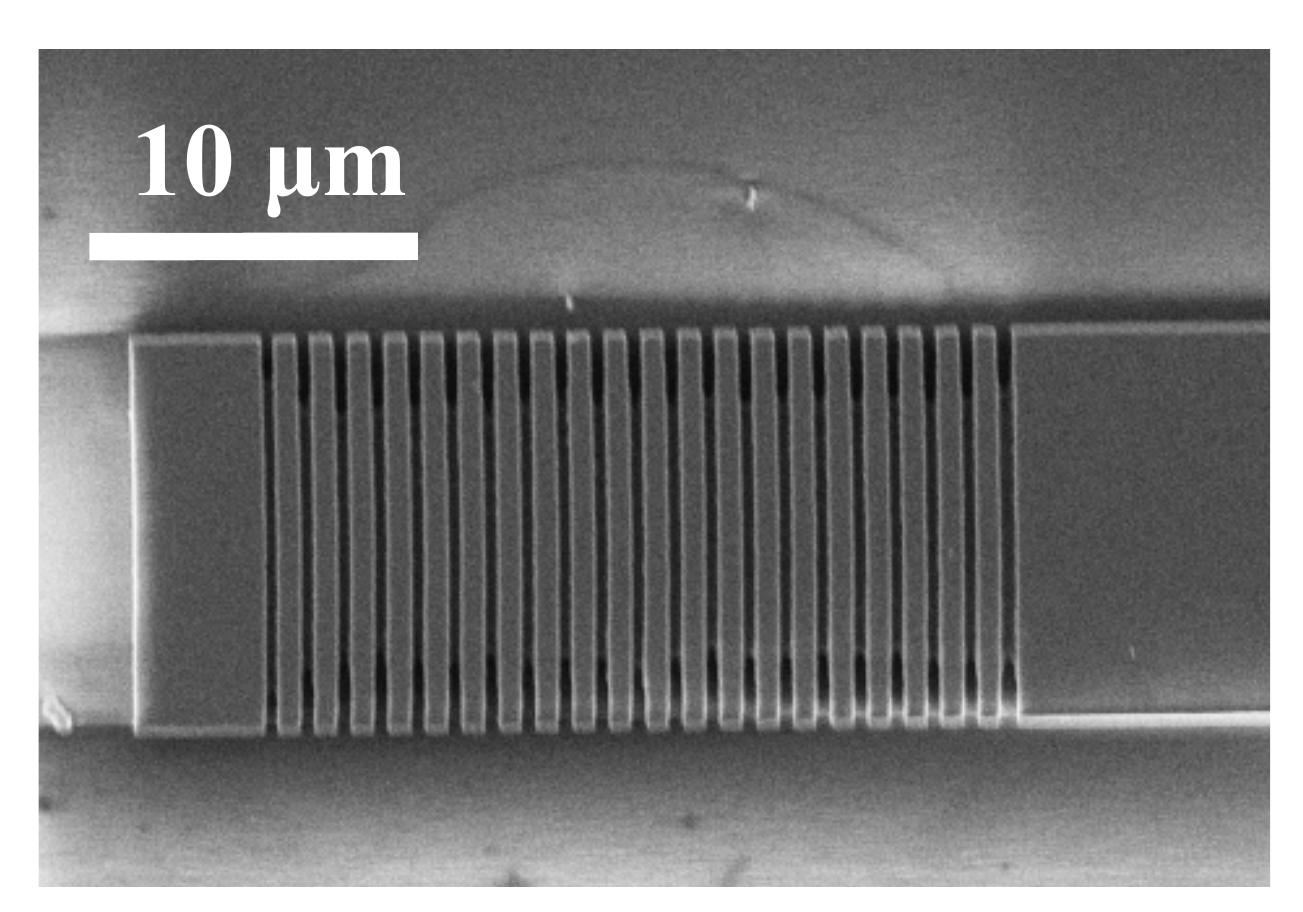}
\caption{$TM$ grating}
\label{fig:subim72}
\end{subfigure}
\begin{subfigure}{0.33\textwidth}
\centering\includegraphics[width=3.8cm]{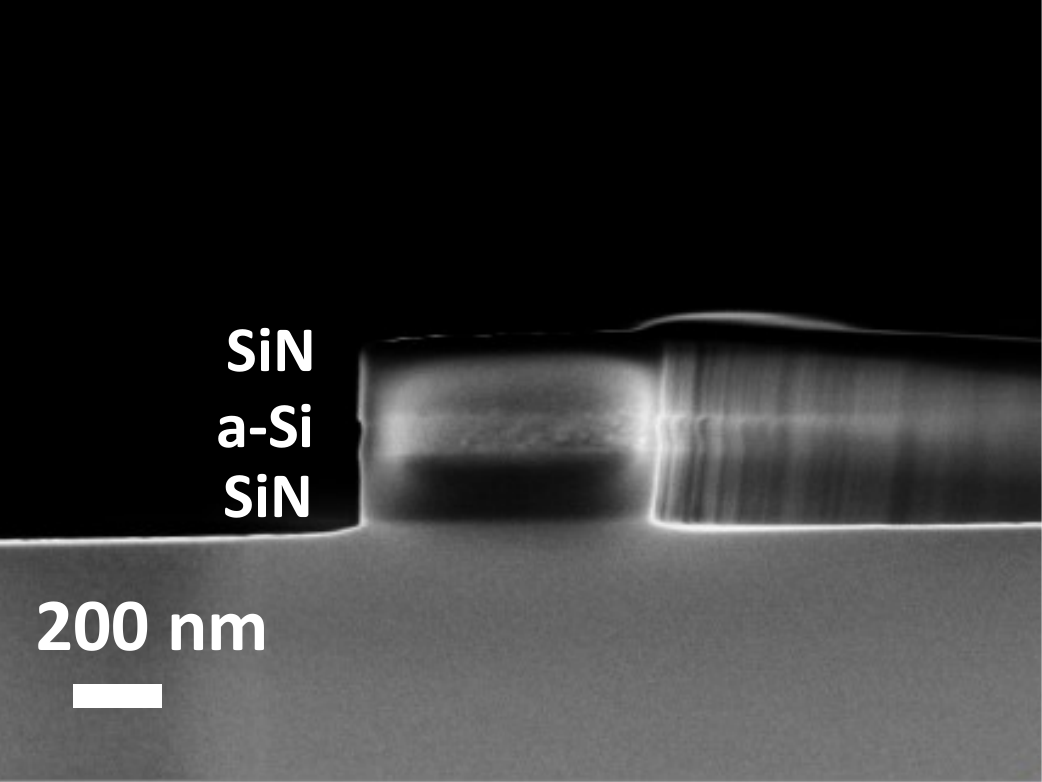}
\caption{Waveguide cross-section}
\label{fig:subim73}
\end{subfigure}

\caption{ SEM image of a fabricated (a) TE grating in $a-Si$ layer, (b) top view of full stack etched $TM$ grating, and (c) cross-section of 700 nm wide and 20\% fraction fabricated waveguide showing the three layers.}

\label{fig:image3}
\end{figure}

The grating designs are fabricated in the desired layer stack. The stack is prepared by depositing $SiO_{2}$, $a-Si$ and $SiN$ using plasma-enhanced chemical vapour depositing process. The grating and waveguide patterning is done using the electron-beam lithography process. Figure \ref{fig:im6} shows the process flow for fabrication of both $TE$ and $TM$ gratings and the waveguide. The entire stack is deposited for $TM$ mode, followed by lithography and pattern transfer of both gratings and waveguide through etching. In $TE$ mode, the gratings are patterned in the $a-Si$ layer, followed by $SiN$ deposition. After that, the waveguide is patterned on the completed stack. Figures \ref{fig:subim71} and \ref{fig:subim72} show the top view of the $TE$ and $TM$ gratings respectively. Figure \ref{fig:subim73} shows the cross-section of a 700 nm wide waveguide with three layers.

The TE and TM gratings were made in different runs with a common oxide layer. The deposited oxide thickness is 1570 nm. The deposited thickness of SiN layer is 220 nm for TE gratings and 214 nm for TM gratings. The deposited thickness of a-Si layer is 98 nm for TE gratings and 115 nm for TM gratings. The etching of SiN layers was done with $CHF_3$ and $O_2$ and a-Si was done with $SF_6$ and $CHF_3$.

The grating test structures are made on a 12 $\mu$m wide waveguide with \textbf {} number of periods. The fabricated devices are characterised using a superluminescent diode (SLED) as a source and a spectrum analyzer. The input polarisation is controlled by polarization paddles. The coupling efficiency is calculated by normalizing the transmission to the source spectrum. 

\begin{figure}[t!]
\begin{subfigure}{0.33\textwidth}
\centering\includegraphics[width=4.3cm]{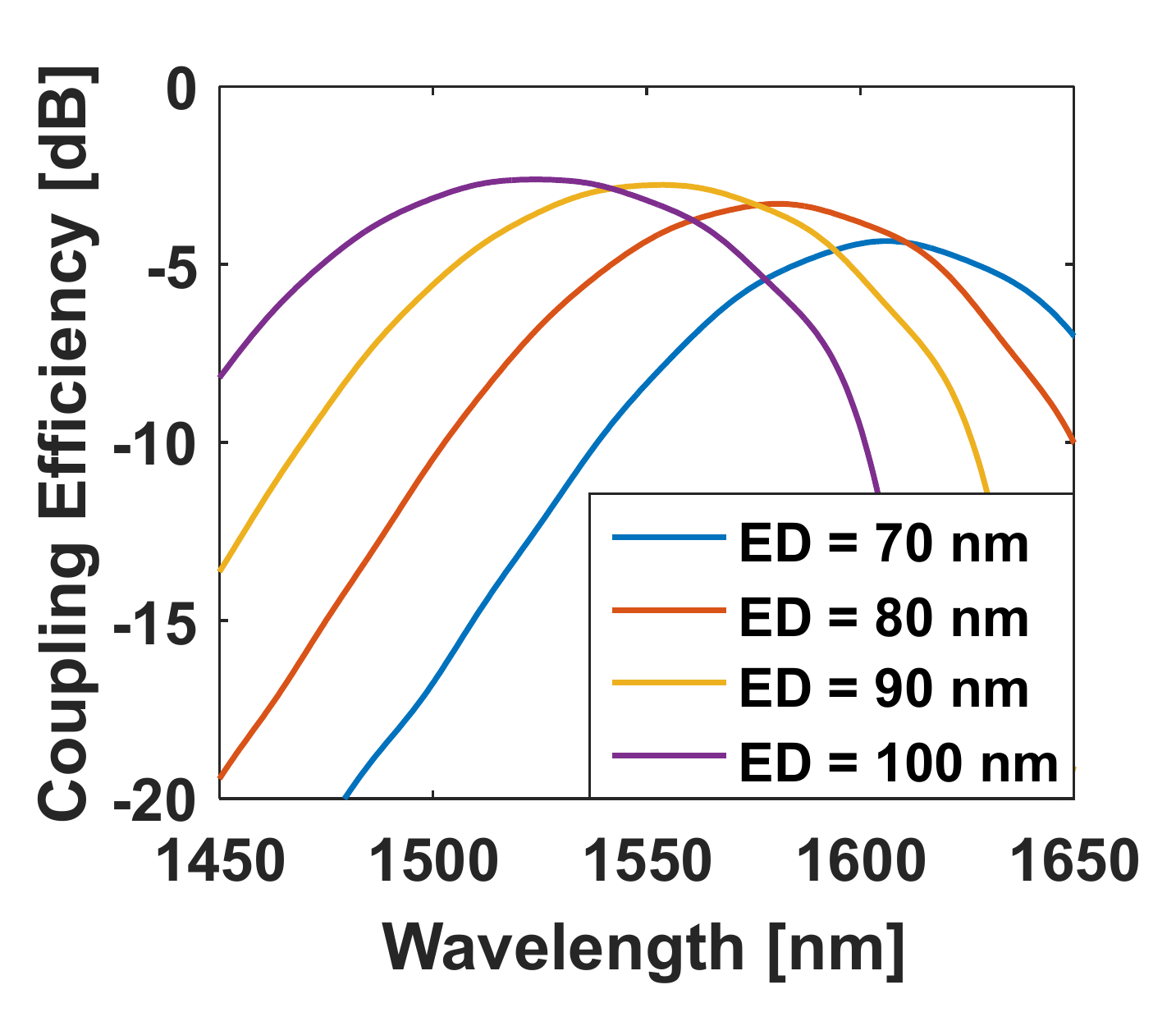}
\caption{CE vs ED (Simulated )} 
\label{fig:subim81}
\end{subfigure}
\begin{subfigure}{0.32\textwidth}
\centering\includegraphics[width=4.3cm]{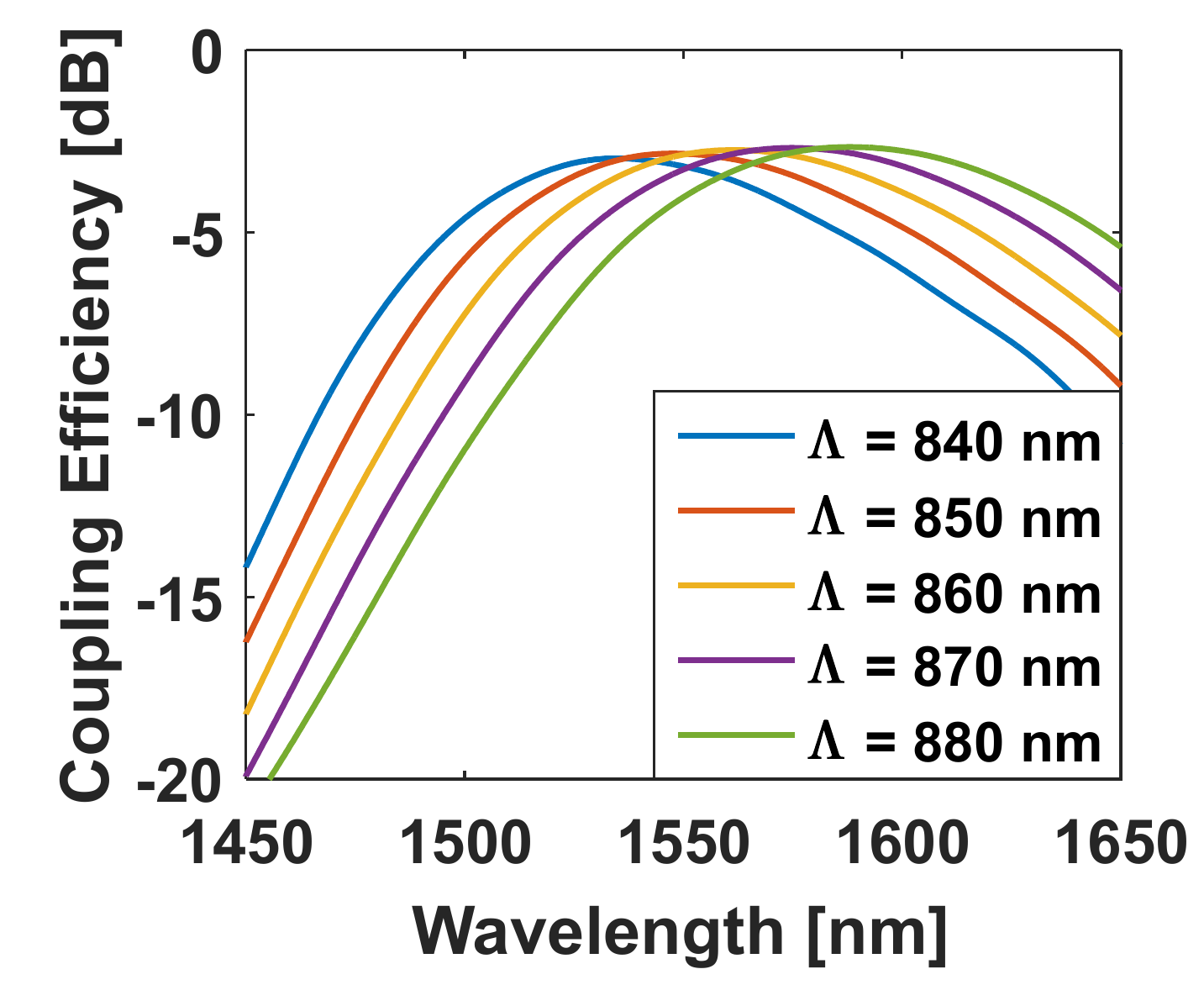}
\caption{CE vs period (Simulated)}
\label{fig:subim82}
\end{subfigure}
\begin{subfigure}{0.33\textwidth}
\centering\includegraphics[width=4.3cm]{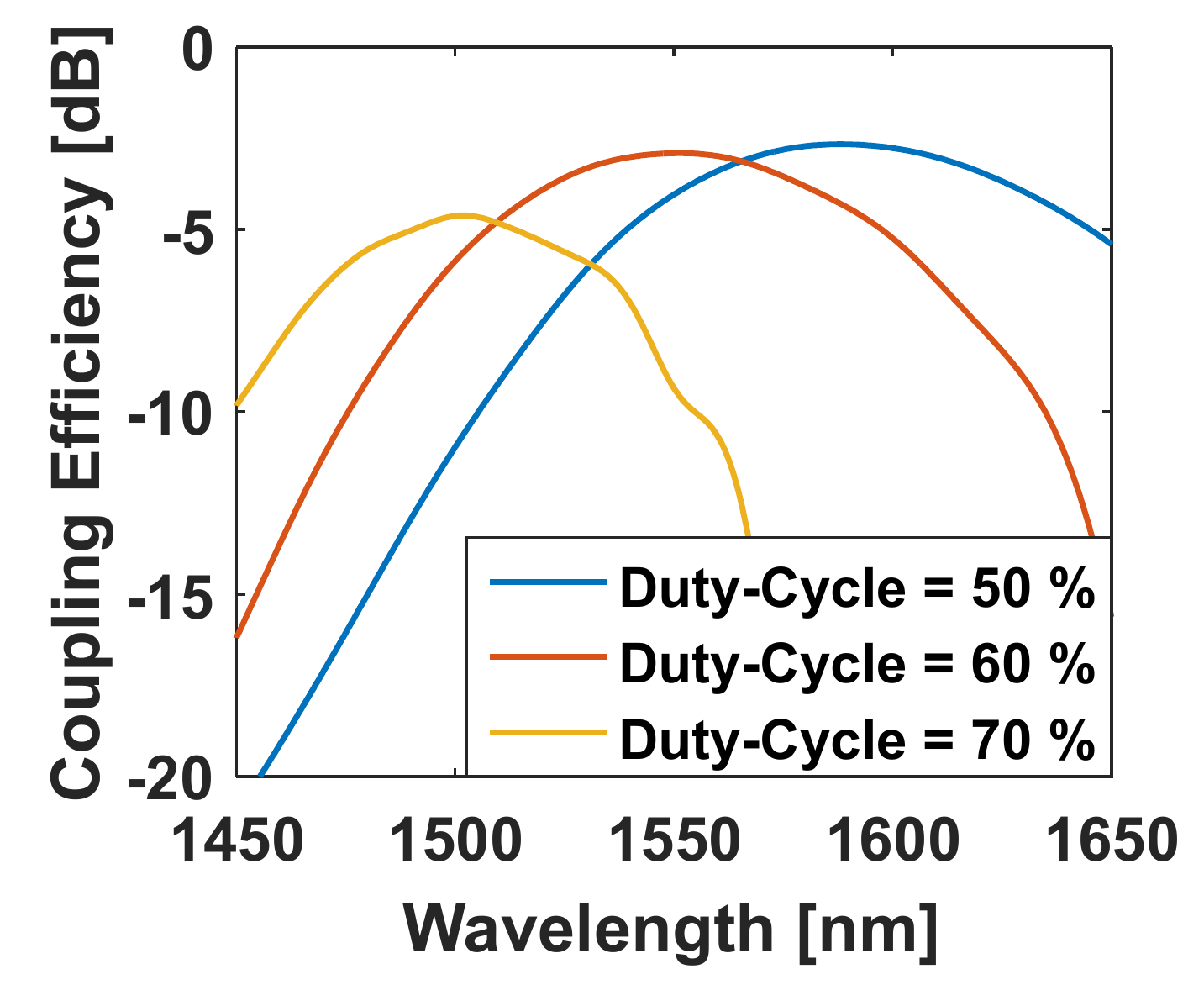}
\caption{CE vs duty-cycle (Simulated)}
\label{fig:subim83}
\end{subfigure}
\begin{subfigure}{0.33\textwidth}
\centering\includegraphics[width=4.3cm]{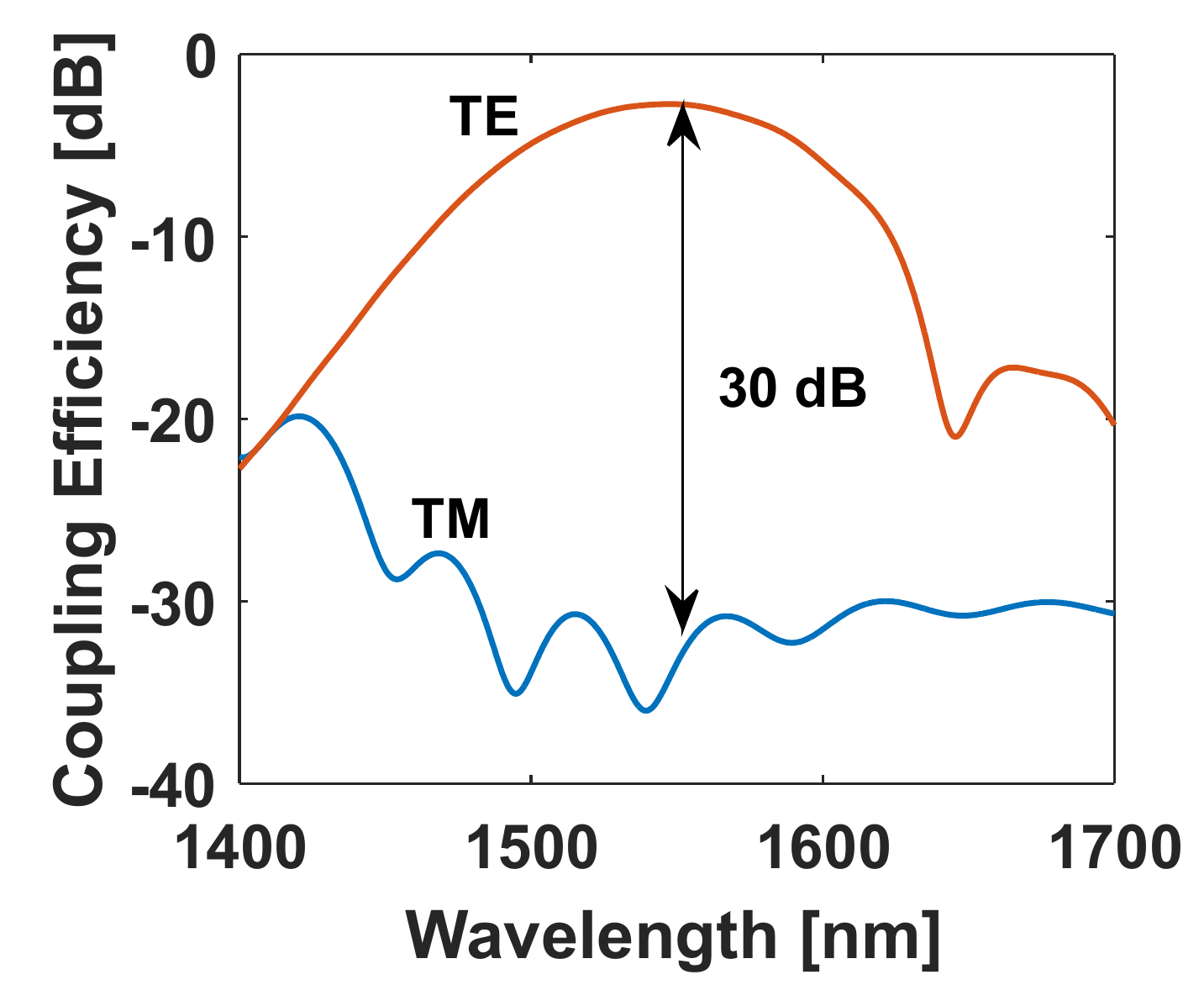}
\caption{$TE$ and $TM$ extinction (simulated) }
\label{fig:subim84}
\end{subfigure}
\begin{subfigure}{0.32\textwidth}
\centering\includegraphics[width=4.3cm]{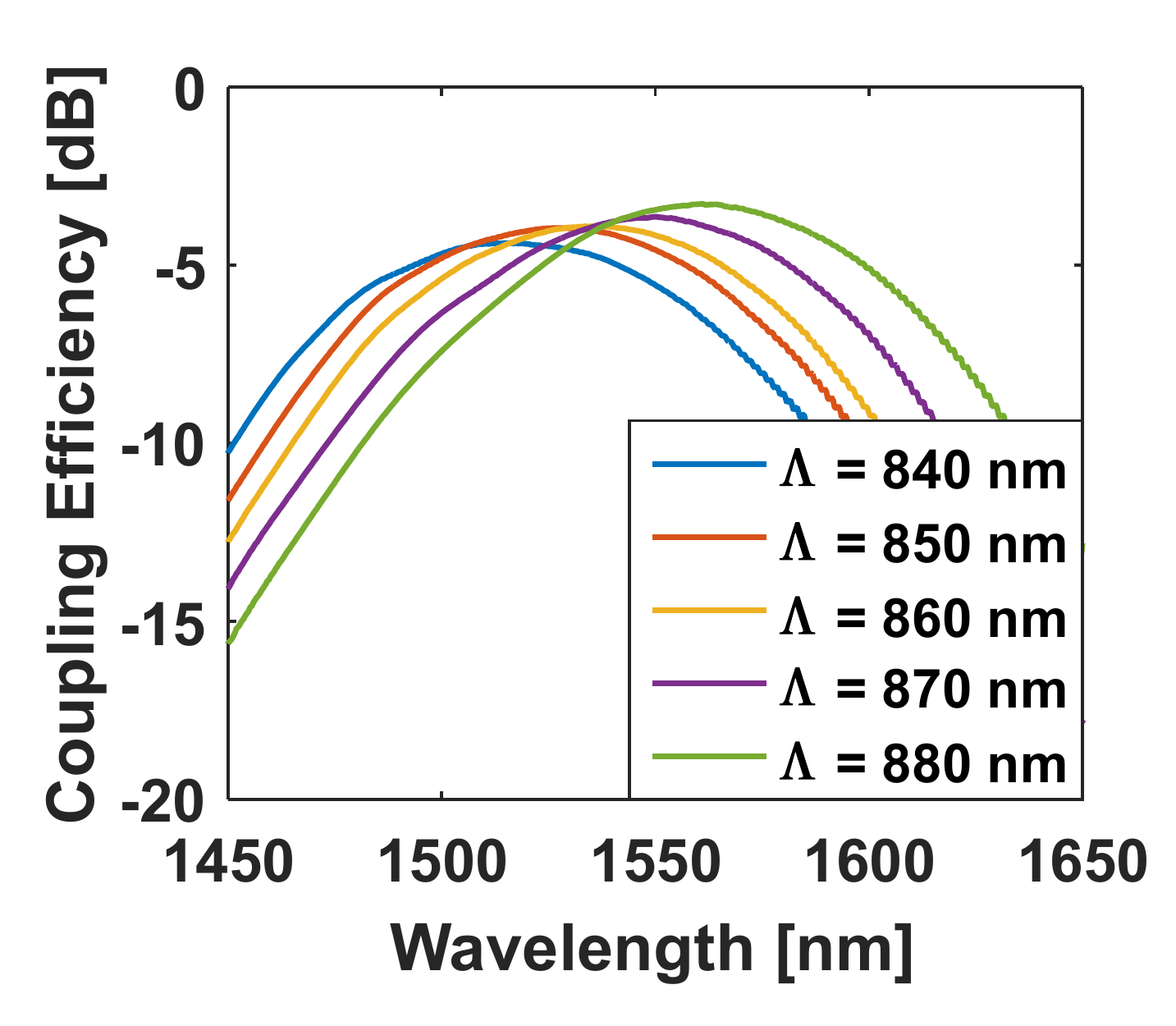}
\caption{CE vs period (Measured)}
\label{fig:subim85}
\end{subfigure}
\begin{subfigure}{0.33\textwidth}
\centering\includegraphics[width=4.3cm]{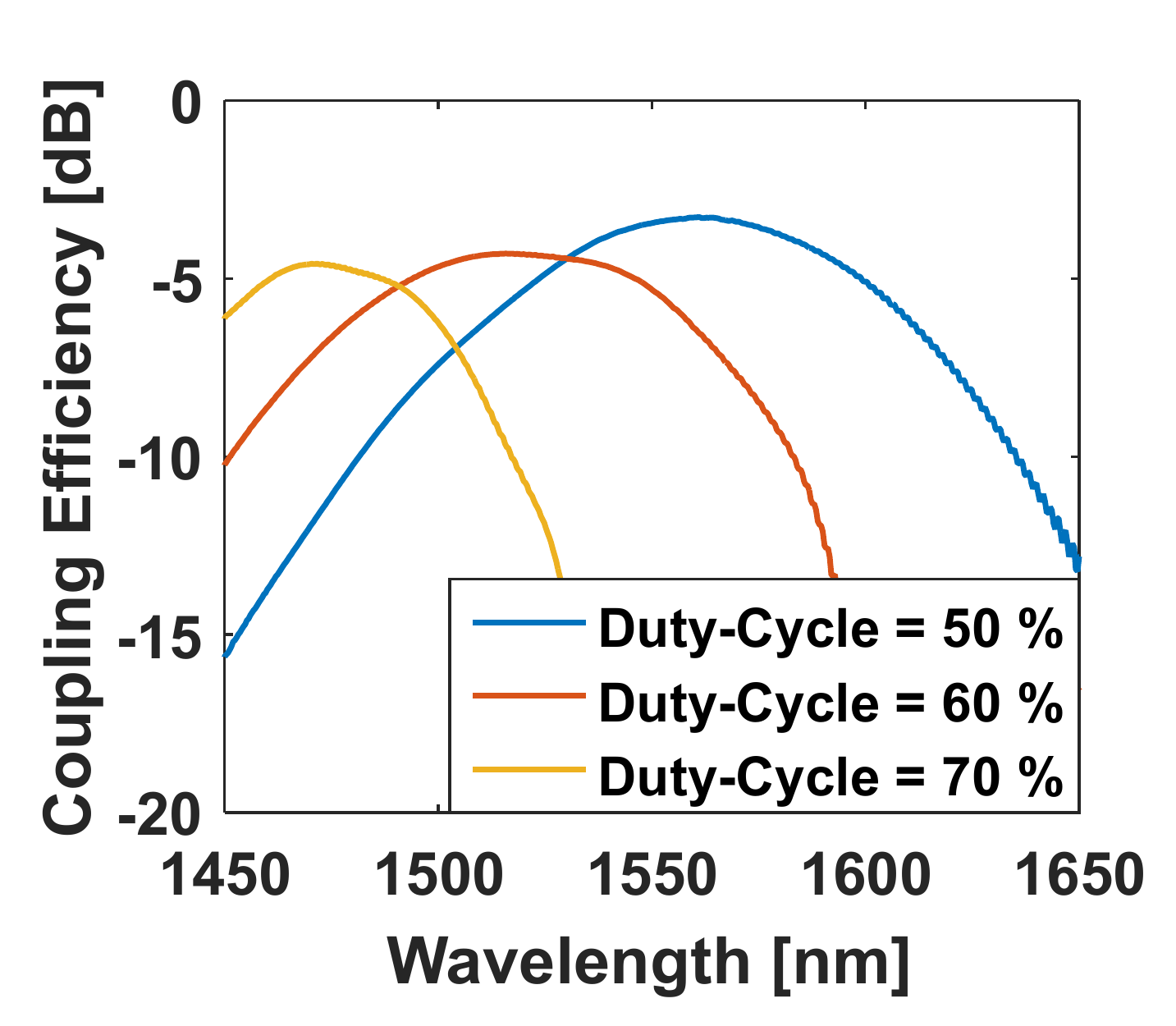}
\caption{CE vs duty-cycle (Measured)}
\label{fig:subim86}
\end{subfigure}

\caption{Summary of simulation and measurement of $TE$ grating couplers. Simulated coupling efficiency (CE) of a $TE$ gratings to (a) etch depth of in 100 nm $a-Si$, (b) grating period, and (c) grating duty-cycle. (d) Extinction of $TM$ and $TE$ of a $TE$ gratings. (e) Measured spectral response of $TE$ gratings as a function of etch depth. (f) Measured spectral response of $TE$ gratings as a function of duty-cycle. }
\label{fig:im8}
\end{figure}

\begin{figure}[ht!]

\begin{subfigure}{0.33\textwidth}
\centering\includegraphics[width=4.3cm]{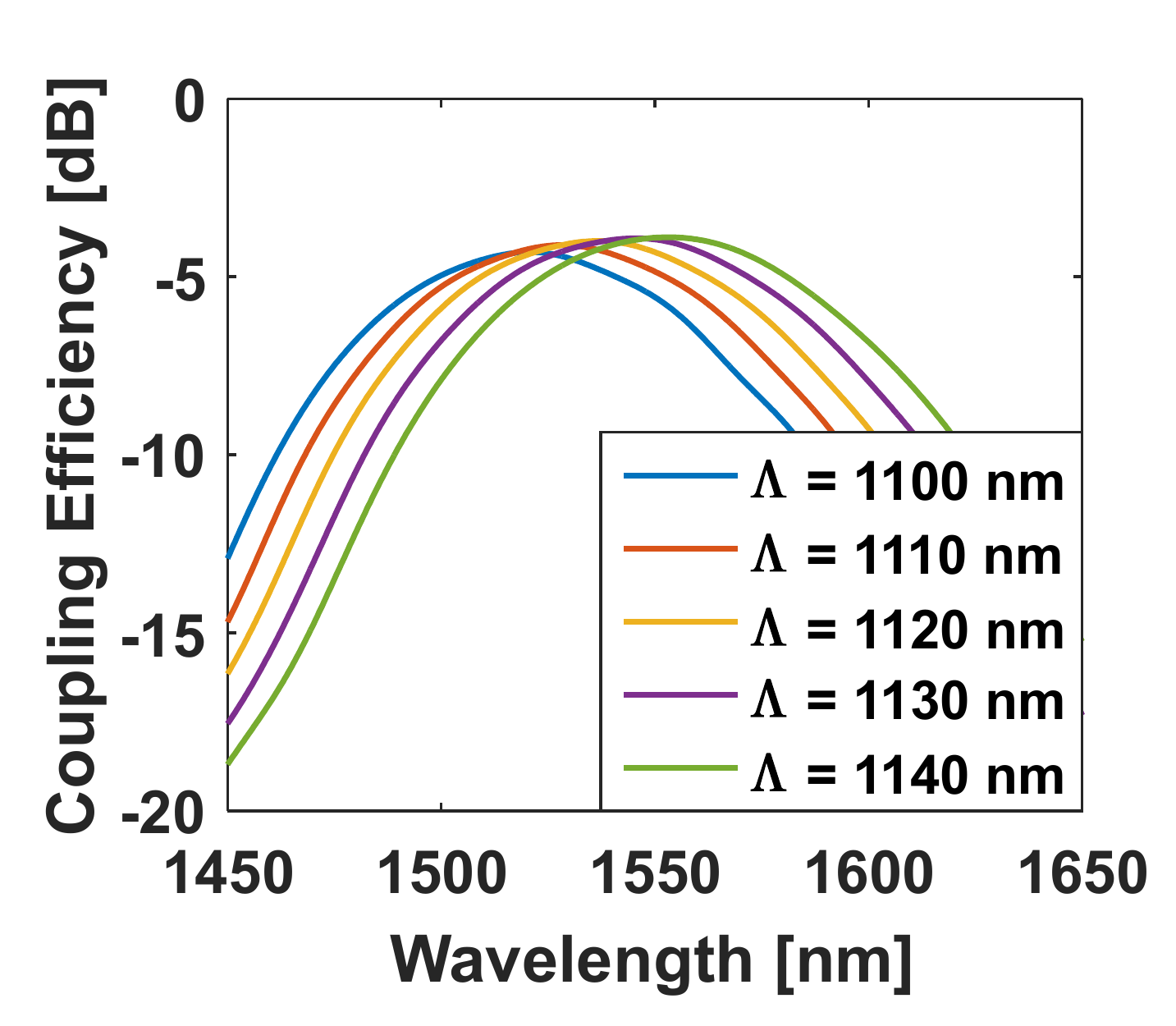}
\caption{CE vs period (Simulated TM)}
\label{fig:subim101}
\end{subfigure}
\begin{subfigure}{0.32\textwidth}
\centering\includegraphics[width=4.3cm]{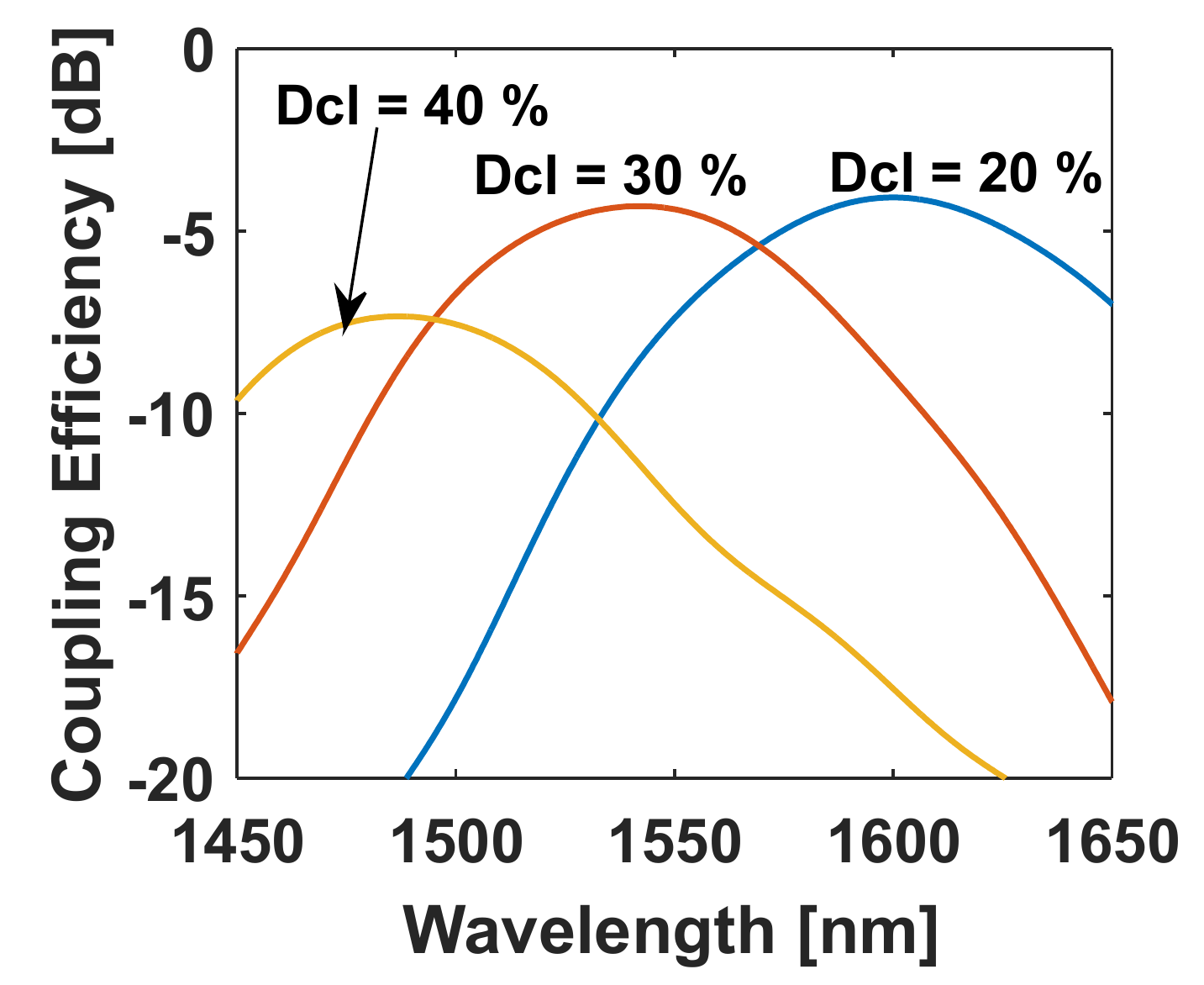}
\caption{CE vs duty-cycle (Simulated TM)}
\label{fig:subim102}
\end{subfigure}
\begin{subfigure}{0.33\textwidth}
\centering\includegraphics[width=4.3cm]{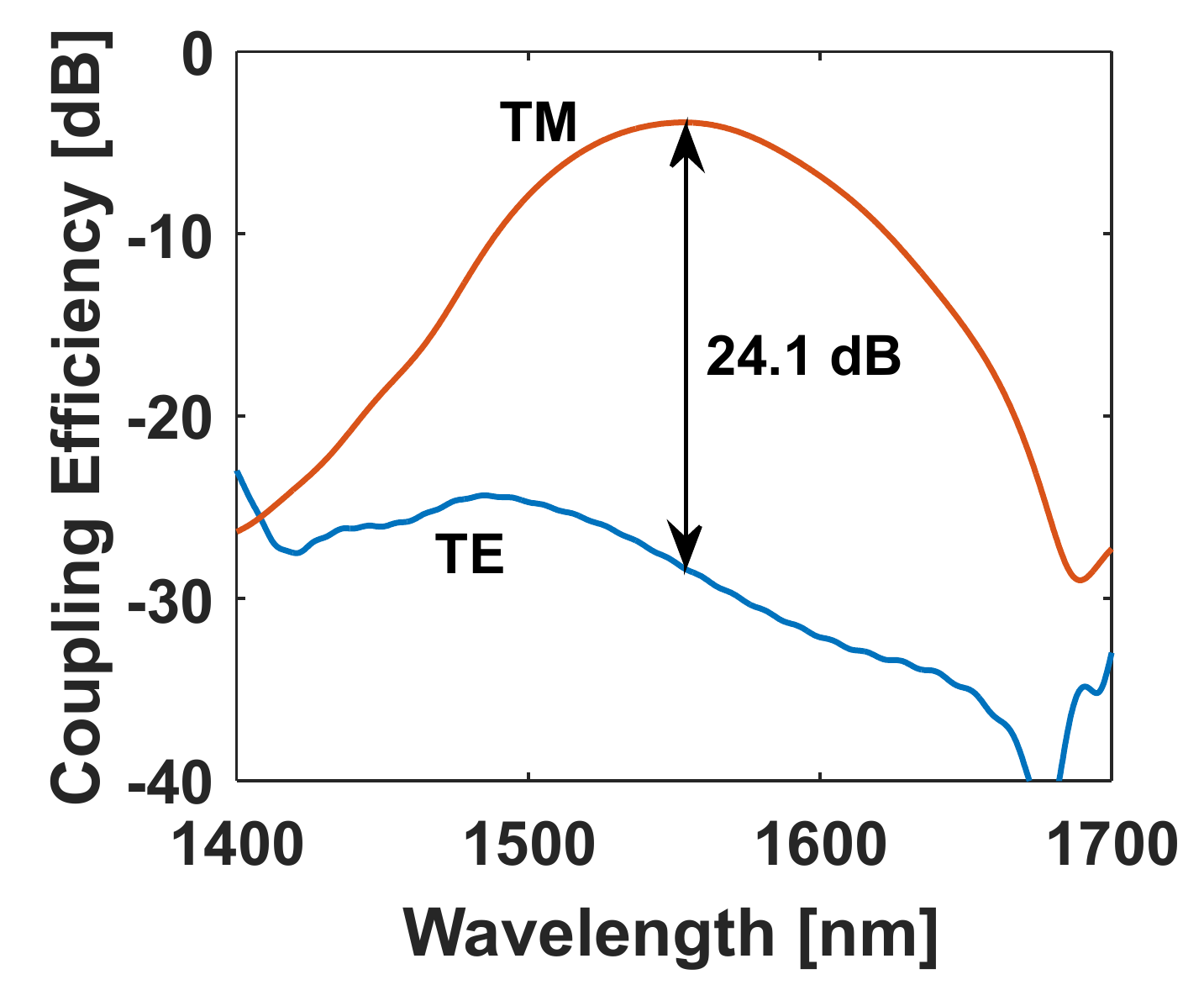}
\caption{$TE$ and $TM$ extinction (simulated)}
\label{fig:subim103}
\end{subfigure}
\begin{subfigure}{0.33\textwidth}
\centering\includegraphics[width=4.3cm]{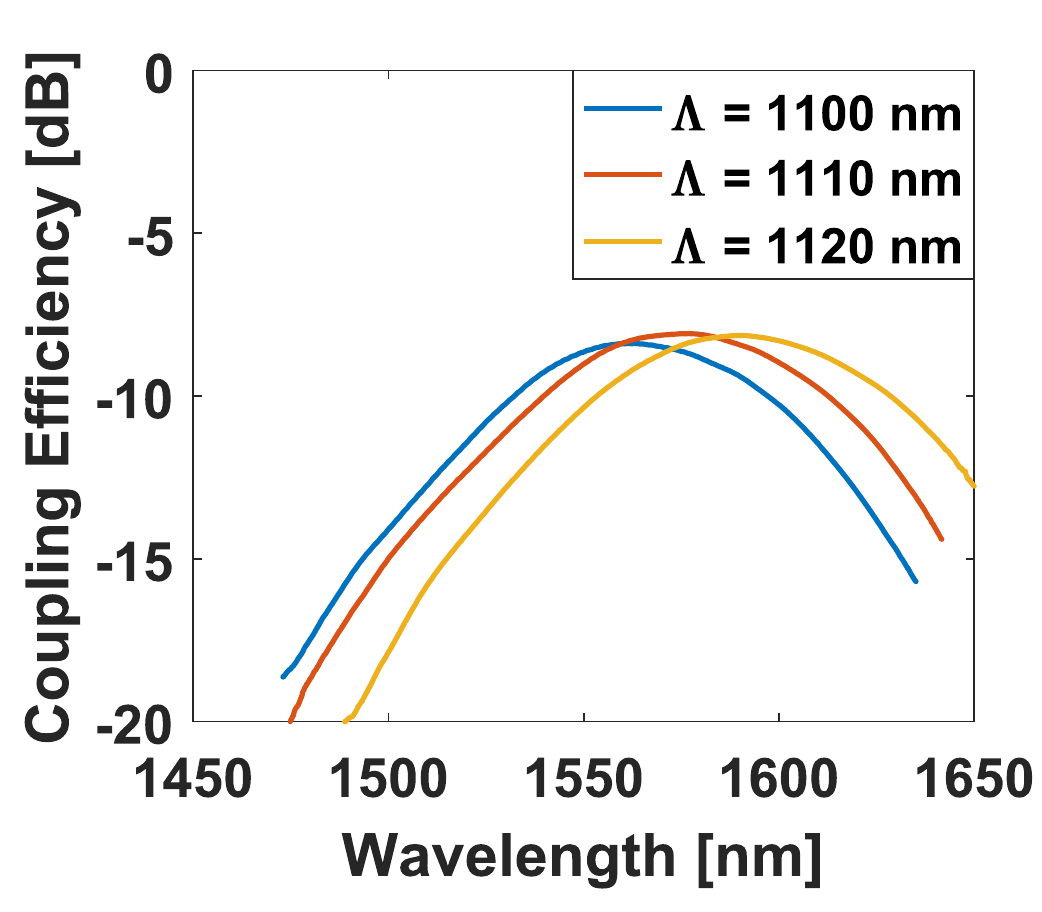}
\caption{CE vs period (Measured TM)}
\label{fig:subim104}
\end{subfigure}
\begin{subfigure}{0.32\textwidth}
\centering\includegraphics[width=4.3cm]{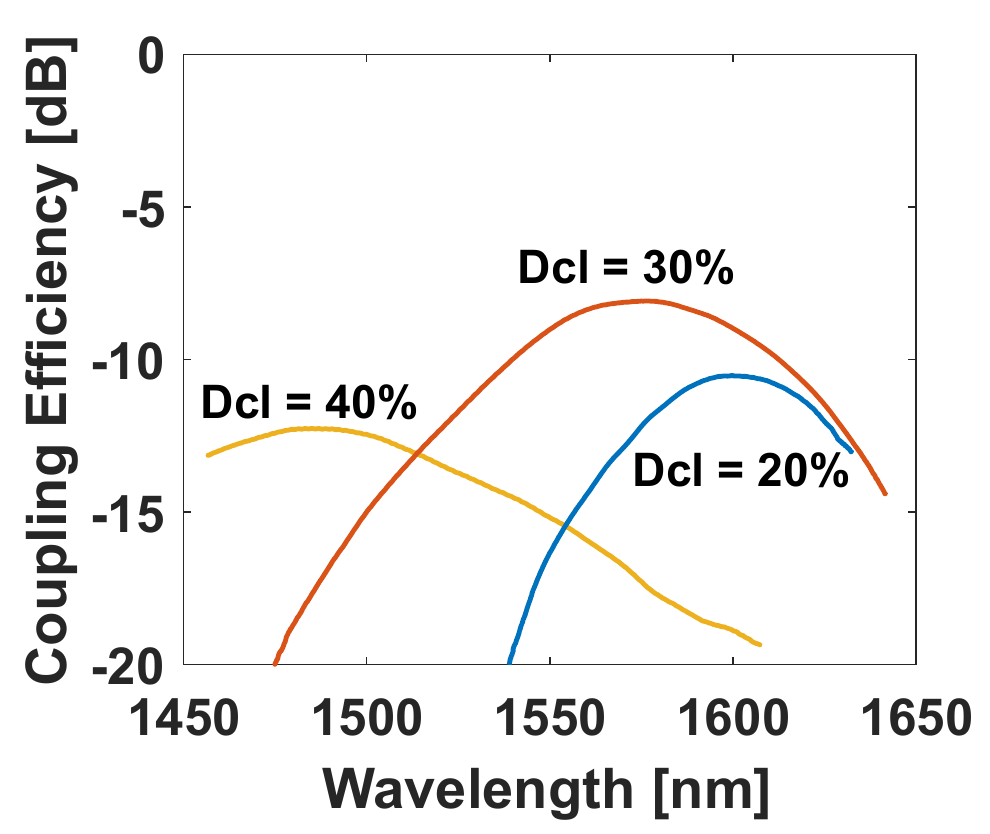}
\caption{CE vs duty-cycle (Measured TM)}
\label{fig:subim105}
\end{subfigure}
\begin{subfigure}{0.33\textwidth}
\centering\includegraphics[width=4.3cm]{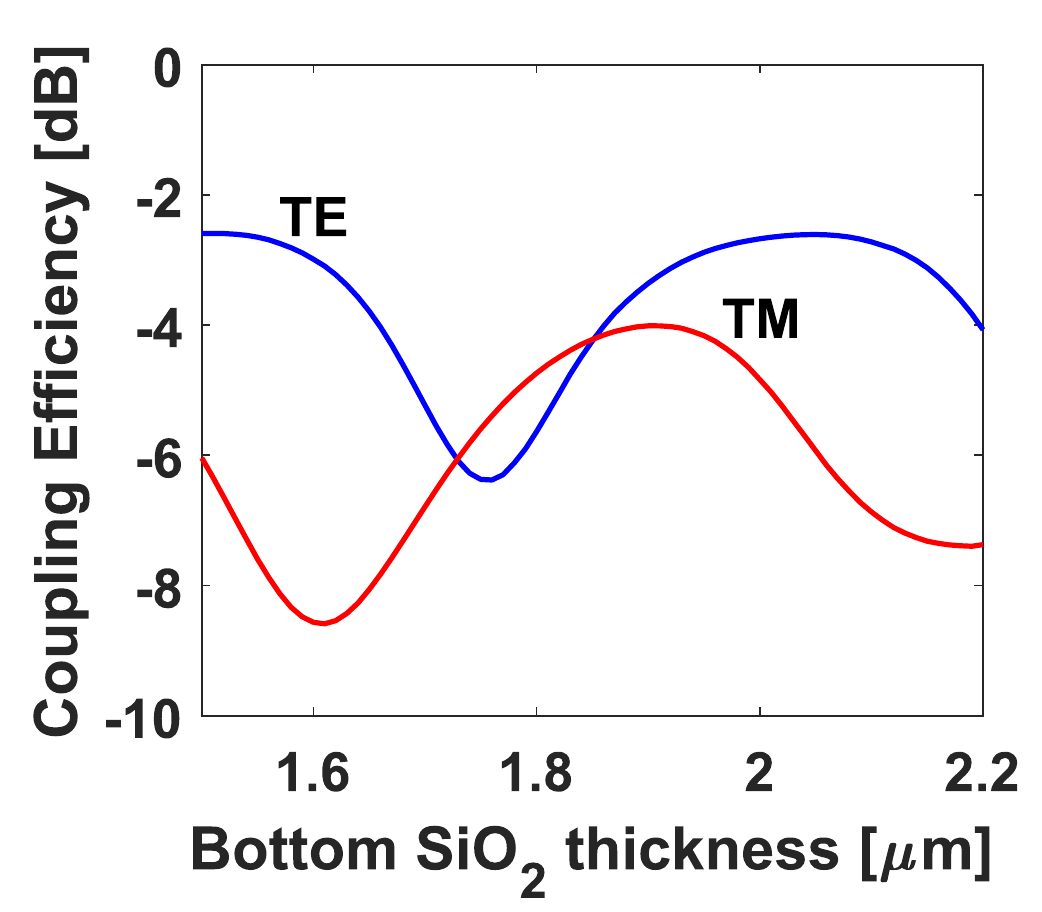}
\caption{}
\label{fig:subim106}
\end{subfigure}

\caption{Summary of simulation and measurement of $TM$ grating couplers. Simulated coupling efficiency (CE) of a $TM$ gratings to (a) grating period and (b) duty-cycle. (c) Extinction of $TM$ and $TE$ of a $TM$ gratings. Measured Coupling efficiency of $TM$ gratings as a function of (d) etch depth and (e) duty-cycle. (f) Effect of bottom oxide thickness on the CE of $TE$ and $TM$ gratings.}
\label{fig:im10}
\end{figure}

\subsection{Results and analysis}

Figure \ref{fig:im8} depicts the summary of simulation and measured spectral response of $TE$ grating coupler. Figures \ref{fig:subim81}, \ref{fig:subim82} and \ref{fig:subim83} show the effect of etch depth (ED), period and duty-cycle, respectively, on the coupling efficiency of a $TE$ grating coupler. It is known that the 1D gratings are polarization sensitive, Fig. \ref{fig:subim84} depicts the simulated extinction between $TE$ and $TM$ input on $TE$ gratings. The experimental validation is depicted in Figures \ref{fig:subim85} and \ref{fig:subim86} showing the measured spectral response of $TE$ gratings with varying period and duty-cycle. The Bragg's law can be used to verify the coupler characteristics. The grating period is related to the coupling wavelength ($\lambda$), fibre angle ($\theta$) and effective index ($n_{eff}$) as, 

\begin{equation}\label{eq3}
\Lambda = \frac{\lambda}{n_{eff}-n_c\sin(\theta).}
\end{equation}

We observe that there is a red shift of the spectrum with an increase in period, both in simulation and experiment. This can be explained from Eq. (\ref{eq3}), where the centre wavelength is directly proportional to period. With an increase in duty-cycle, the spectrum is blue shifted for both simulation and experimental measurements and can be verified from Eq. (\ref{eq3}). The increase in duty-cycle decreases the effective index resulting in blue shift. The etch depth increase also results in blue shift due to decrease in effective index. The extinction for $TM$ polarized light is about 30 dB with respect to $TE$ polarized light in $TE$ mode gratings.

Similar analysis is done for $TM$ mode gratings. Figures \ref{fig:subim101} and \ref{fig:subim102} show the simulated variation of spectral response of $TM$ gratings with period and duty-cycle. Figure \ref{fig:subim103} shows the simulated extinction between $TE$ and $TM$ input on $TM$ gratings. Figures \ref{fig:subim104} and \ref{fig:subim105} show the measured variation of spectral response of $TM$ gratings with period and duty-cycle.

\begin{figure}[h!]
\begin{subfigure}{0.5\textwidth}
\centering\includegraphics[width=4.5cm]{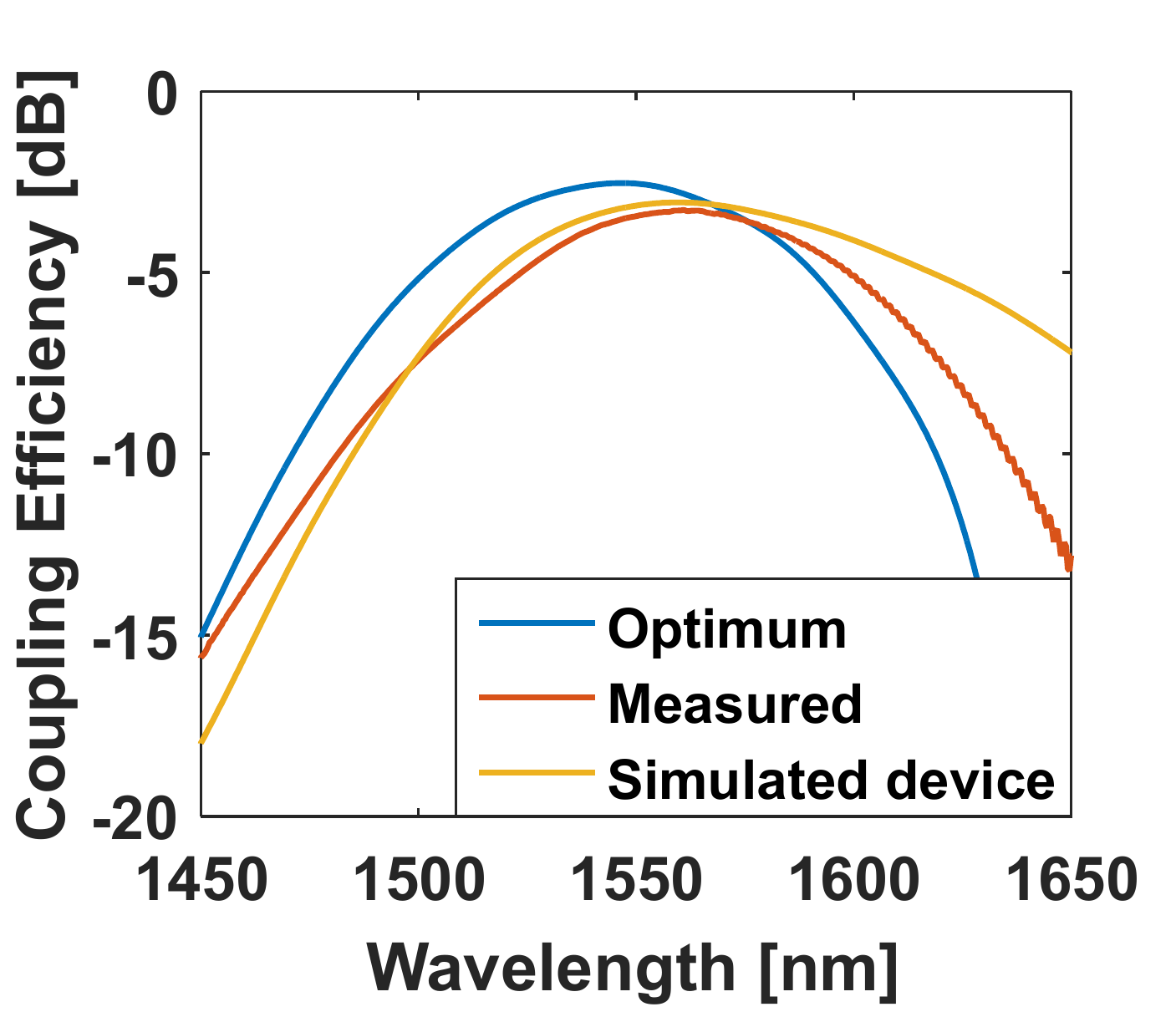}
\caption{$TE$ grating}
\label{fig:subim121}
\end{subfigure}
\begin{subfigure}{0.3\textwidth}
\centering\includegraphics[width=4.5cm]{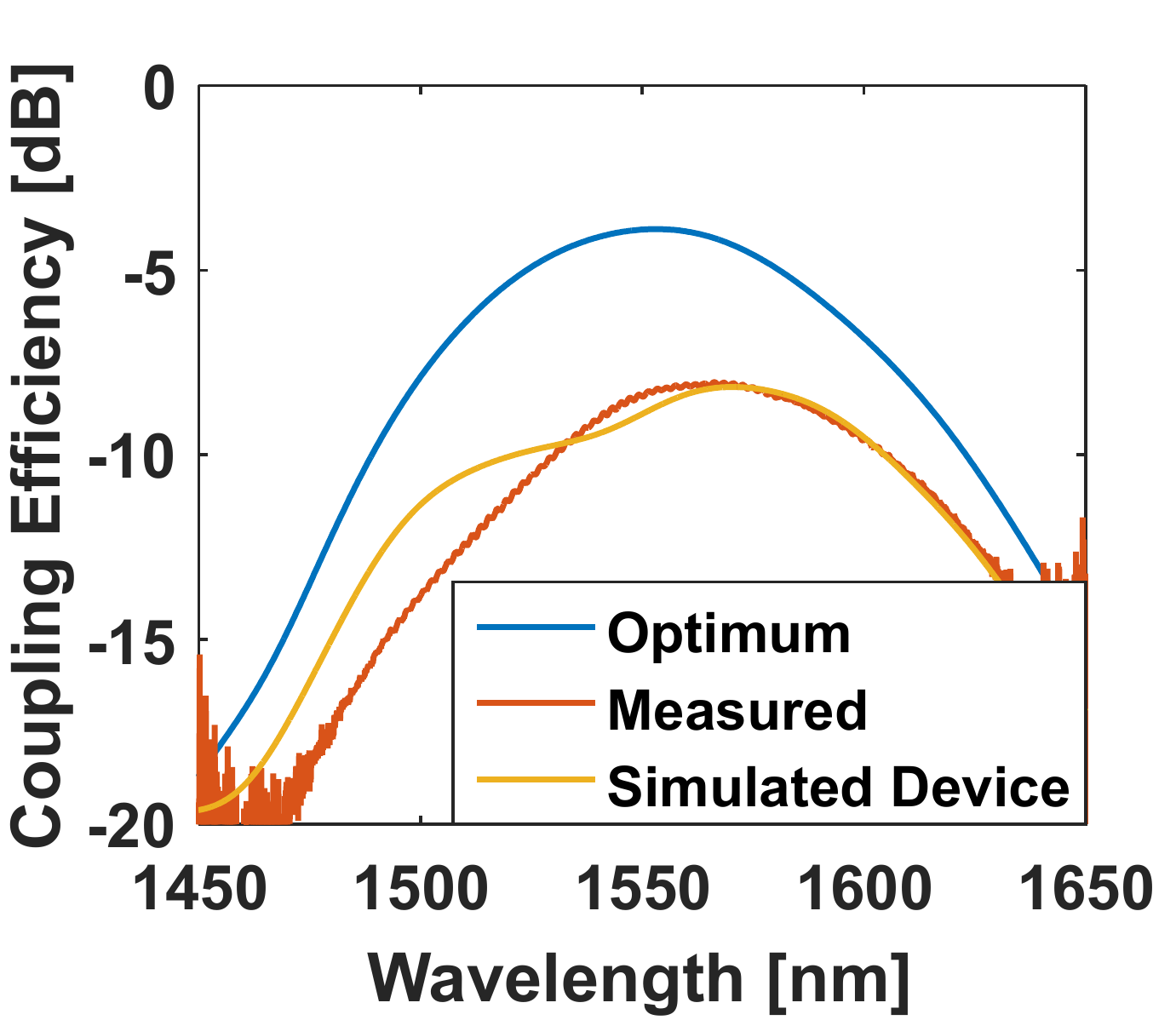}
\caption{$TM$ grating}
\label{fig:subim122}
\end{subfigure}

\caption{ Optimal design and measured coupling efficiency of (a) $TE$ and (b) $TM$ grating couplers. }
\label{fig:im12}
\end{figure}

\begin{table}[ht!]
\caption{Comparison of measured and optimized grating response}
\centering\begin{tabular}{|p{1.9cm}||p{1cm}|c|p{0.8cm}|p{0.8cm}||p{1cm}|c|p{0.8cm}|p{0.8cm}|}
\hline
&\multicolumn{4}{c||}{  $TE$ }& \multicolumn{4}{c|}{  $TM$ } \\
\hline
 Data & CE [dB] & $\lambda_P$[nm] & 1 dB BW [nm] & 3 dB BW [nm]& CE [dB] & $\lambda_P$[nm] & 1 dB BW [nm] & 3 dB BW [nm]\\
 \hline
  Optimum design &	-2.52 &	1547& 57.6 & 98 & -3.9 & 1553 & 54.4 & 94\\
 Fabricated/ Measured &	-3.27 &	1561 & 56.8 &	100 & -8 & 1572 & 51 & 92\\
Fabricated device simulation &	-3.06 &	1560 & 70 &	127 & -8.16 & 1570 & 50 & 113\\

\hline
\end{tabular}
\label{table:table2}
\end{table}
    
The best measured coupling efficiency for $TE$ gratings is -3.27 dB/coupler and for $TM$ gratings is -8 dB per coupler. Figure \ref{fig:subim121} and \ref{fig:subim122} show the comparative spectral response of the optimally designed, fabricated device and measured device for $TE$ and $TM$ mode, respectively. Table \ref{table:table2} summarizes measurement and simulation grating performance. 

The $TE$ grating coupler efficiency is close to the optimum design. However, a red shift in the spectral response from the optimized value is due to a slightly lower than desired etch depth. The coupling efficiency for $TM$ mode is $\approx$4 dB less than optimum. This can be attributed to the SiO$_2$ thickness variation as shown in Figure \ref{fig:subim106} \cite{Sychugov1996}. The energy coupling efficiency can be further increased by employing a bottom Bragg reflector\cite{Selvaraja2009}.

\section{Confirmation of $TE$ and $TM$ mode}
\subsection{Ring Resonator in the hybrid platform}

\begin{figure}[htbp]
\begin{subfigure}{0.5\textwidth}
\centering\includegraphics[width=6cm]{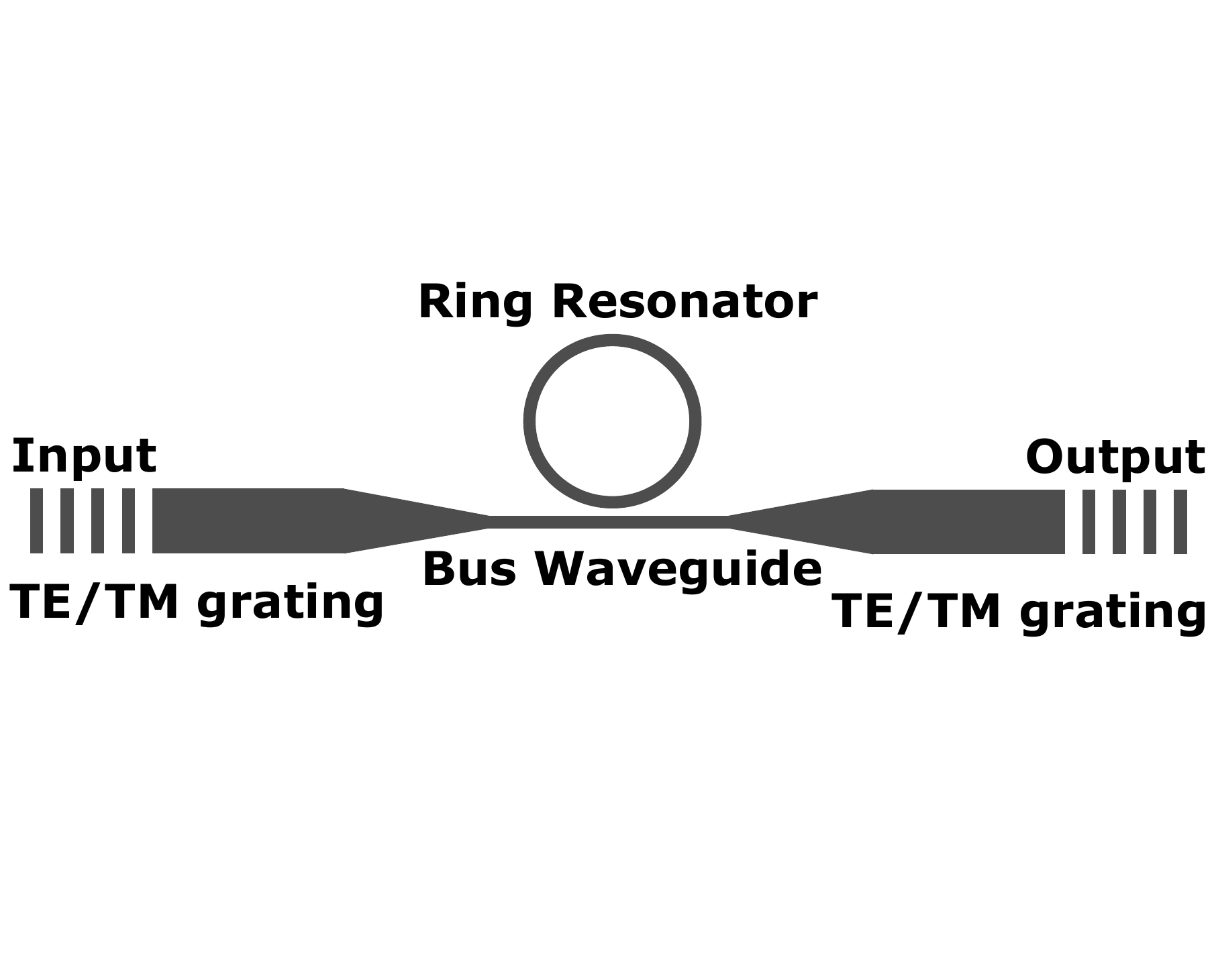}
\caption{}
\label{subim131}
\end{subfigure}
\begin{subfigure}{0.5\textwidth}
\centering\includegraphics[width=6cm]{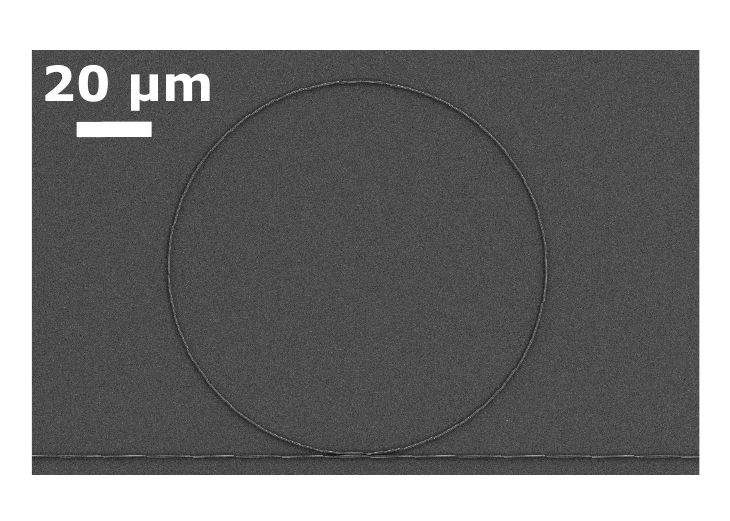}
\caption{}
\label{subim132}
\end{subfigure}

\caption{(a) Schematic of the test structure and (b) SEM of the fabricated ring resonator}
\label{fig:Ring_schematic_sem}
\end{figure}

The propagation of the desired mode in the waveguide can be characterized by using a wavelength selective device, such as a ring resonator. The spectral characteristics of a ring resonator depend on the group index of resonant mode in the cavity. By calculating the group index from the free-spectral range (FSR), one could identify the cavity mode polarization \cite{Bogaerts2012}. The above-mentioned stack is used to fabricate a 100 $\mu$m diameter ring resonator. The waveguide width is kept at 700 nm to accommodate both $TE_{0}$ and $TM_{0}$ (Fig. \ref{im2}). A gap of 250 nm and 350 nm is used between the bus and the ring for $TE$ and $TM$ rings. Figure \ref{subim131} shows the schematic and SEM image of a device. Waveguide tapers of length 250 $\mu$m are used to connect the 700 nm single-mode waveguide to the grating fiber-couplers made on 12 $\mu$m width waveguide. The in and out couplers are designed to be either $TE$ or $TM$ to selectively excite single polarization in the waveguide. 

\begin{figure}[ht!]

\begin{subfigure}{0.5\textwidth}
\centering\includegraphics[width=6.5cm]{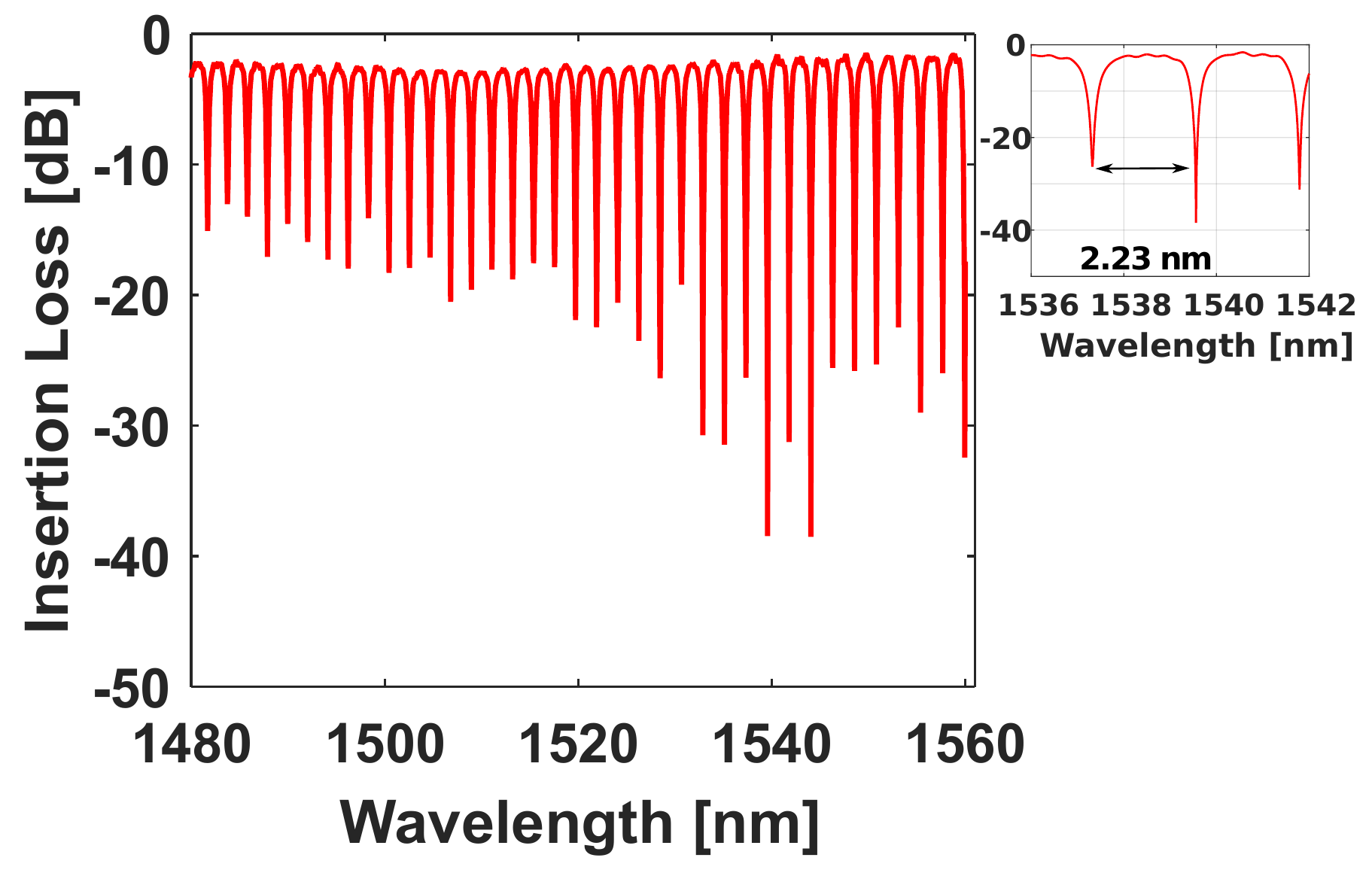}
\caption{$TE$ Ring}
\label{subim141}
\end{subfigure}
\begin{subfigure}{0.5\textwidth}
\includegraphics[width=6.5cm]{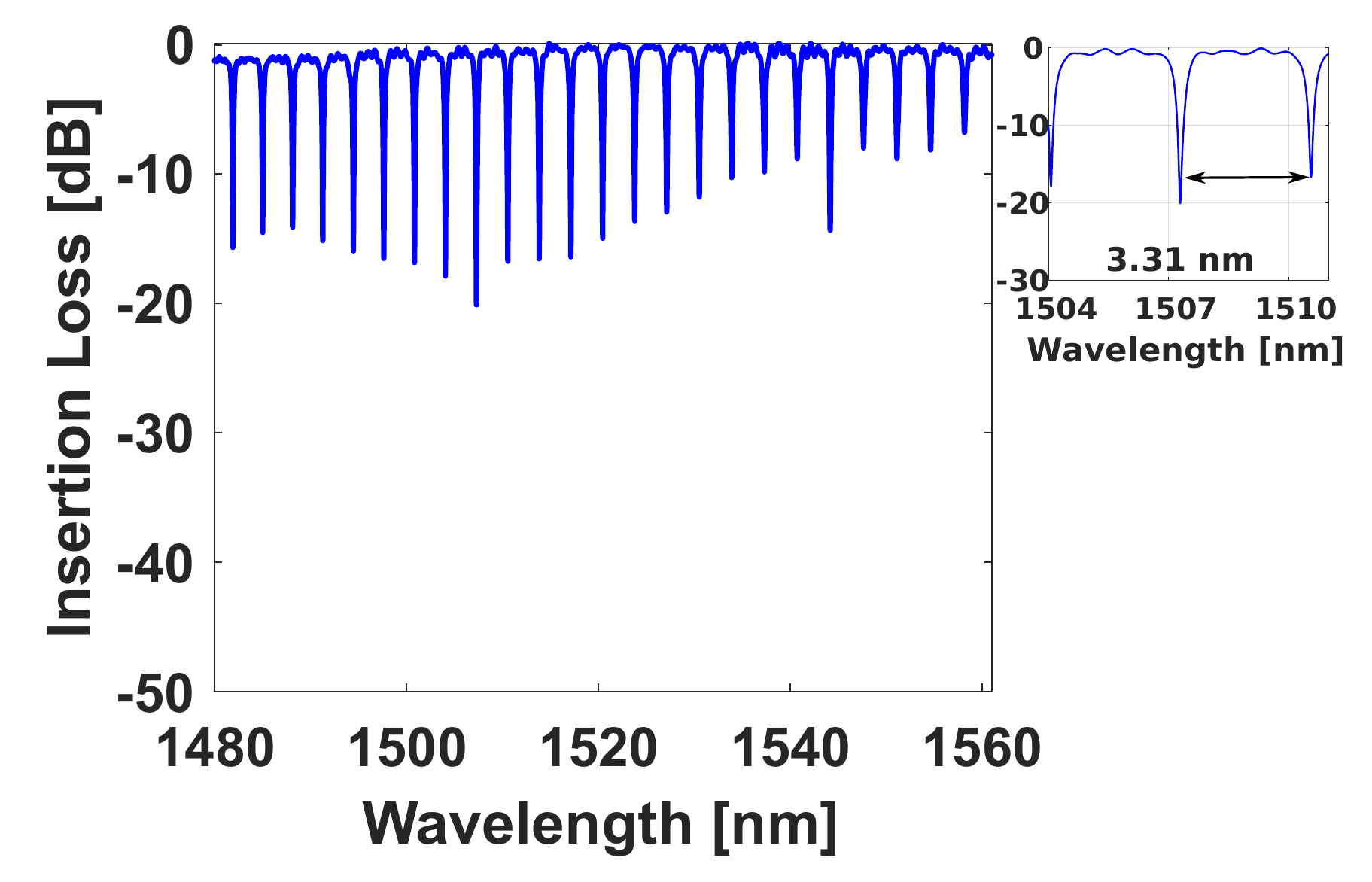} 
\caption{$TM$ Ring}
\label{subim142}
\end{subfigure}

\caption{Normalized Spectral response of (a) $TE$  and (b) $TM$ ring resonators. }
\label{fig:ring_response}
\end{figure}

Figures \ref{subim141} and \ref{subim142} show the normalized spectral response of the $TE$ and $TM$ ring, respectively. An FSR of 2.23 nm and 3.31 nm is measured for $TE$ and $TM$ resonators. The corresponding calculated group index is 3.38 and 2.18 against an expected 3.48 and 2.33 for $TE$ and $TM$, respectively. This shows an excellent agreement and confirmation of corresponding polarization excitation by the grating couplers. In addition, we observe a quality factor of 3400 for the $TE$ ring and 4000 for the $TM$ ring. The $TM$ resonance shows a better quality factor due to lower scattering from the etched sidewalls. The loss is primarily due to the interface between the sandwich layers. This is further confirmed by a lower insertion loss of about -1 dB from the $TM$ ring compared to -2.2 dB from a $TE$ ring resonator.

\subsection{Thermal response and mode confinement.}

\begin{figure}[ht!]
\begin{subfigure}{0.33\textwidth}
\centering\includegraphics[width=4.5cm]{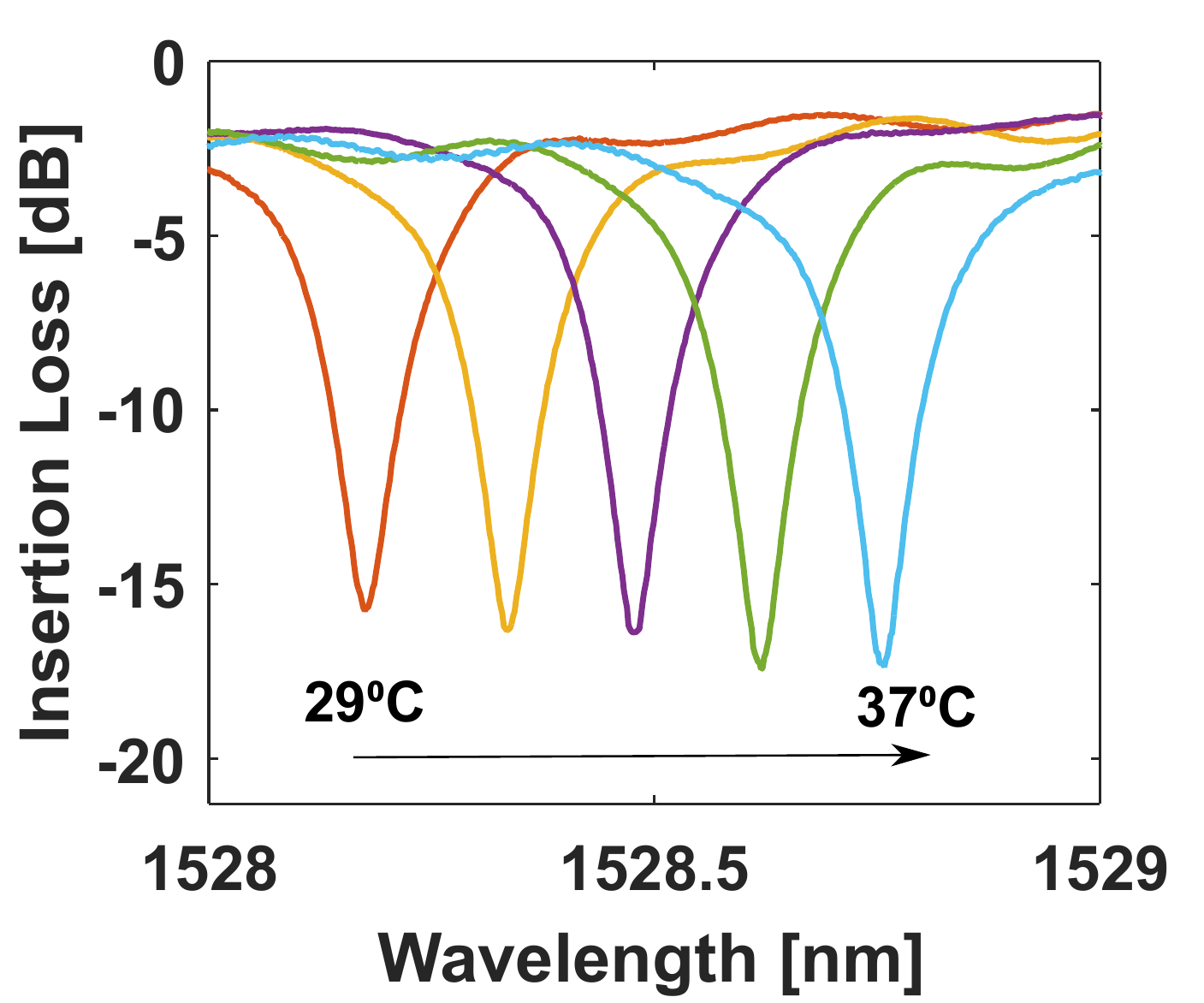}
\centering\caption{}
\label{fig:subim151}
\end{subfigure}
\begin{subfigure}{0.3\textwidth}
\includegraphics[width=4.5cm]{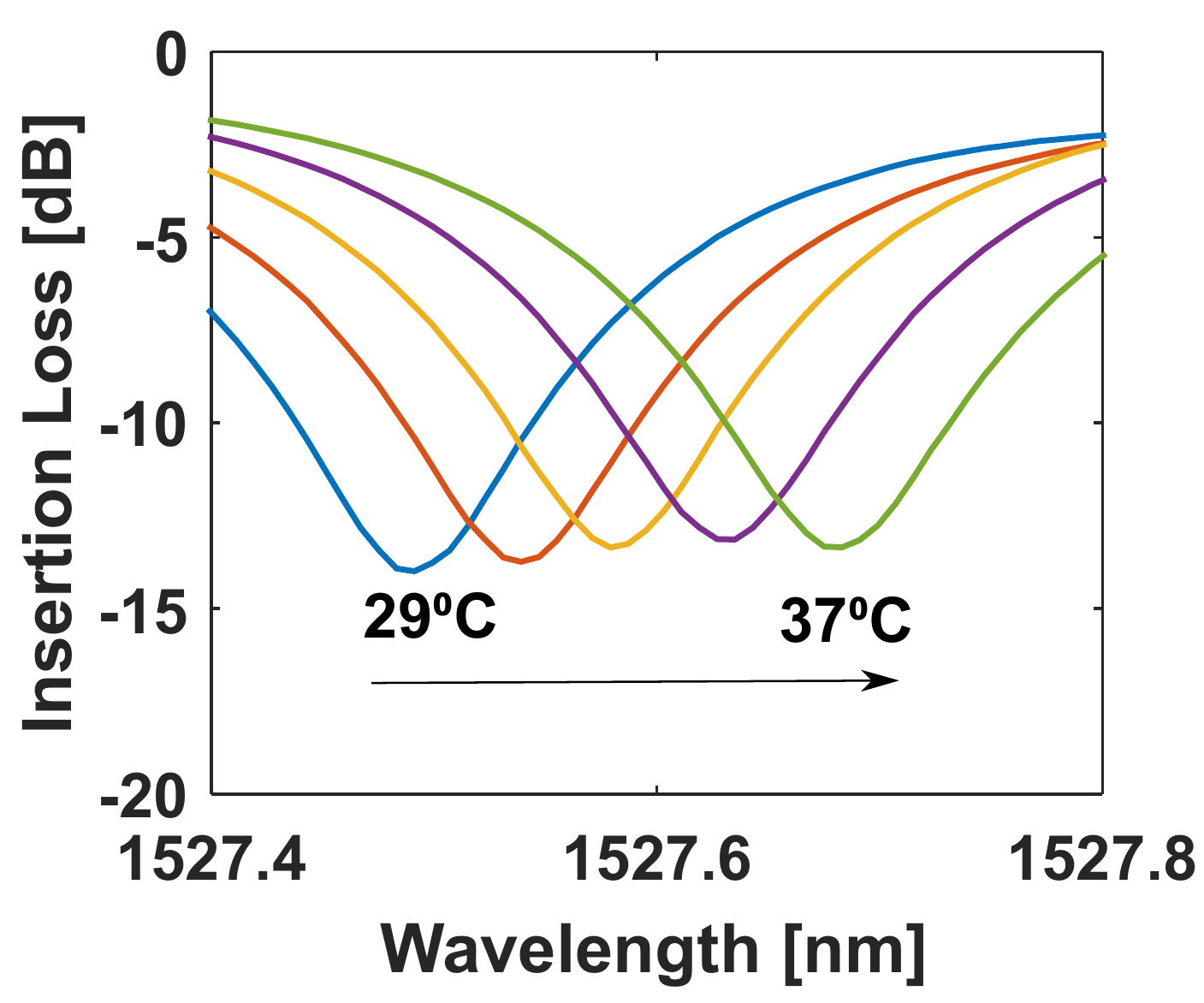}
\caption{}
\label{fig:subim152}
\end{subfigure}
\begin{subfigure}{0.3\textwidth}
\includegraphics[width=4.5cm]{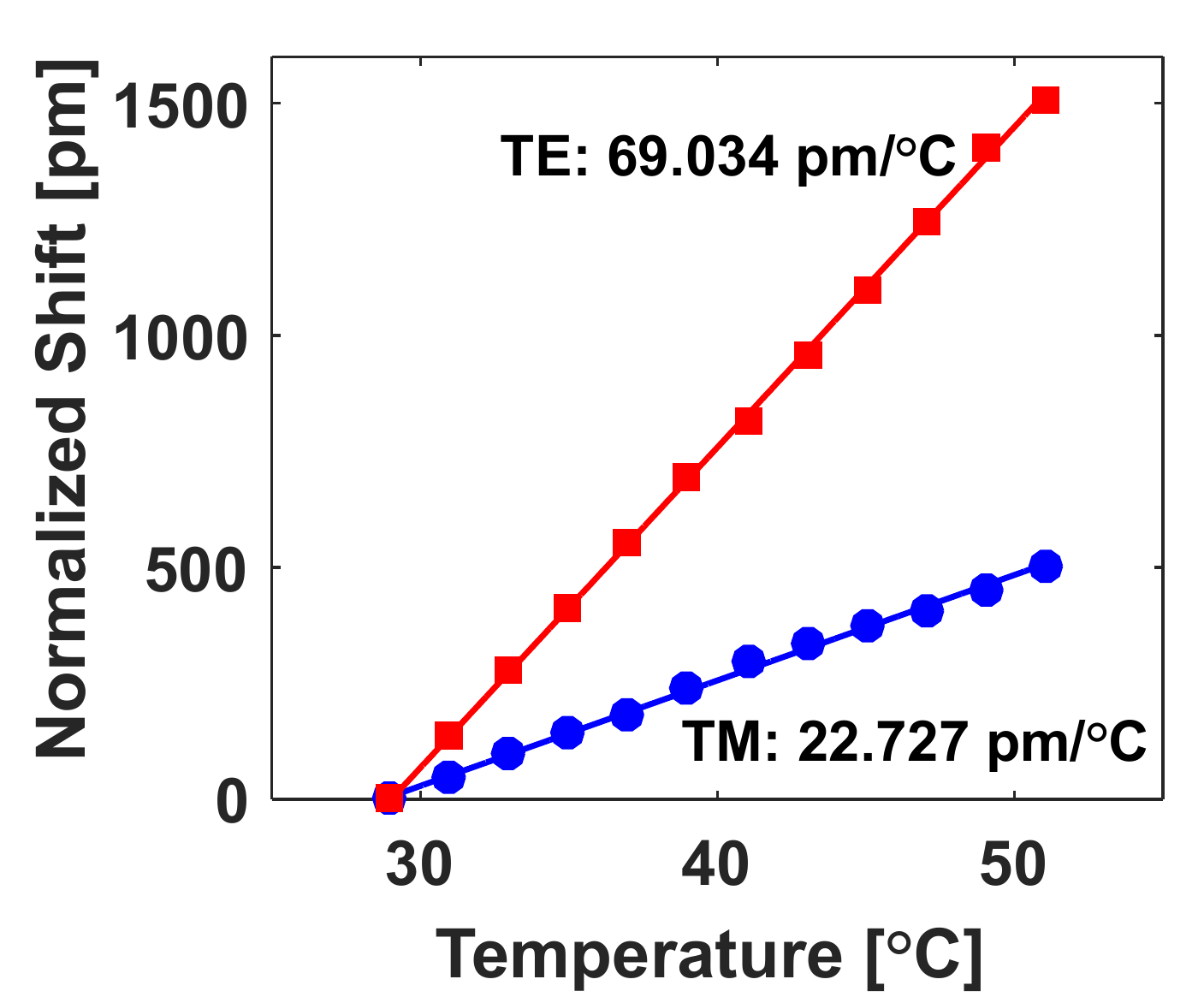}
\centering\caption{}
\label{fig:subim153}
\end{subfigure}
\caption{Thermo-optic response of the hybrid waveguide. Spectral shift with temperature for (a) $TE$ polarization and (b) $TM$ polarization. (c) Linear thermo-optic response shift for $TE$ and $TM$.}
\label{im15}
\end{figure}

Unlike a solid single material core, the mode distribution in the hybrid waveguide is specific to polarization $TE$ and $TM$. As discussed in section \ref{Waveguide_Design}, The $TE_0$ mode is primarily confined in the $a-Si$ while the $TM_0$ is distributed between the two $SiN$ layers. Since the mode confinement is hybrid, the confinement can be confirmed using the thermo-optic response of the two modes \cite{Teng2009, Feng2015, Mere2018}. The thermo-optic response of the above-mentioned rings is characterized by heating the substrate using a Peltier element and a temperature controller. The temperature is tuned from 29 $^{\circ}$C to 51 $^{\circ}$C in steps of 2 $^{\circ}$C. Figures \ref{fig:subim151} and \ref{fig:subim152} show the shift in resonance wavelength with temperature. The resonance shift is 69 pm/$^{\circ}$C for the $TE$ ring and 23 pm/$^{\circ}$C for the $TM$ ring, as shown in Figure \ref{fig:subim153}. The confirmation of the excited mode and confinement is done by deriving the thermo-optic coefficient (TOC) of the individual materials.

The effective TOC is found from the spectral shift (d$\lambda_R$/dT ) as given by \cite{Teng2009}.

\begin{equation}\label{eq4}
\frac{dn_{eff}}{dT} =\frac{d\lambda_R}{dT}\times\frac{n_g}{\lambda_R}
\end{equation}

The effective TOC of the sandwich waveguide is the superposition of TOC of individual layers and the electric field confinement ($\Gamma$) in the particular layer for a given mode by \cite{Feng2015}.

\begin{equation}\label{eq5}
\frac{dn_eff}{dT} = \Sigma_i\Gamma_i\frac{dn_i}{dT} = \Gamma_{SiN}\frac{dn_{SiN}}{dT} + \Gamma_{a-Si}\frac{dn_{a-Si}}{dT}
\end{equation}

The electric field confinement ($\Gamma_i$) in the ith layer is calculated by taking the electric field distribution \cite{Visser1996}.

\begin{equation}\label{eq6}
\Gamma_i = \frac{\int_{i}^{} |E(x)|^2 dx}{\int_{-\infty}^{+\infty} |E(x)|^2 dx}
\end{equation}

Equation \ref{eq5} is solved simultaneously for $TE$ and $TM$ polarizations to find TOC of $a-Si$ and $SiN$. The calculations are summarized in table 3. The TOC for $a-Si$ is found to be $2.05\times10^{-4}$ and for $SiN$ $2.67\times10^{-5}$ which are close to the values found in literature \cite{DellaCorte2001,{Elshaari2016}} confirming the analysis.

\begin{table}[ht!]
\begin{center}
\caption{Summary of TOC calculation}
\begin{tabular}{|c||c|c|}

    \hline
     Parameter [units] & $TE$ mode & $TM$ mode \\
     \hline
    Measured $d\lambda_R/dT[pm/^{\circ}C]$&69.034 &22.727 \\   
	$dn_{eff}/dT[K^{-1}]$ & $1.528\times10^{-4}$ & $3.468\times10^{-5}$\\   
    $\Gamma_{a-Si}$ & 0.685 & 0.076  \\
    $\Gamma_{SiN}$ & 0.46 & 0.718  \\   
    $n_g$ & 3.38 & 2.18\\   

     \hline
\end{tabular}    
\end{center}
\label{table:toc1}
\end{table}

\section{Conclusion}
We have demonstrated a hybrid waveguide platform with a high-index material sandwiched between two medium-index materials. The proposed waveguide configuration offers a versatile platform to tune the effective index, confinement and polarization behaviour. We have demonstrated grating fiber-chip couplers for coupling into both $TE$ and $TM$ modes with a coupling efficiency of -3.27 dB and -8 dB per coupler with a 3dB bandwidth of 100 nm. Using thermo-optic response, we confirm the polarisation-dependent mode confinement in high and medium-index layers in the waveguide. We believe the proposed waveguide configuration would help to realize interesting functionalities that are limited by the single material solid-core waveguide platforms.  

\section{Acknowledgement}
We thank DST-SERB for funding this research. We also acknowledge funding from MHRD, MeitY and DST for supporting facilities at the Centre for Nanoscience and Engineering (CeNSE), Indian Institute of Science, Bangalore.



\bibliographystyle{elsarticle-num}
\bibliography{SW_WG_FC.bib}

\end{document}